\documentclass[a4paper, onecolumn, preprintnumbers, nofootinbib, notitlepage, superscriptaddress]{revtex4-1}
\usepackage{textcase}
\usepackage{titlesec}

\makeatletter 
\def\@hangfrom@section#1#2#3{\@hangfrom{#1#2}#3}
\def\@hangfroms@section#1#2{#1#2}
\makeatother

\makeatletter 
\renewcommand\thesection{\@arabic\c@section}
\renewcommand\thesubsection{\thesection.\@arabic\c@subsection}
\renewcommand{\section}{\@startsection{section}{1}{0mm}
    {\baselineskip}%
    {\baselineskip}{\bfseries\large\raggedright}}
\renewcommand{\subsection}{\@startsection{subsection}{2}{0mm}
    {\baselineskip}%
    {\baselineskip}{\bfseries\raggedright}}

\renewcommand\p@subsection{}

\makeatother

\usepackage[a4paper, hdivide={2.245cm,,2.245cm}, vdivide={1.83cm,,3.8cm}]{geometry}

\usepackage{color}
\usepackage[hyperfootnotes=false,colorlinks,citecolor=magenta]{hyperref}
\usepackage{ulem}
\usepackage{youngtab}
\usepackage[utf8]{inputenc}
\usepackage[english]{babel} 
\usepackage{amsmath}
\usepackage{float}
\usepackage{amssymb, url}

\usepackage{braket}
\usepackage{amsfonts,bbm}
\usepackage{dsfont,tensor}
\usepackage{enumerate}
\makeatletter\AtBeginDocument{\let\@elt\relax}\makeatother
\usepackage{graphicx}
\usepackage{bbm}
\usepackage{array}
\usepackage{booktabs}
\usepackage{units}
\usepackage{textcomp}
\usepackage{mathtools}
\usepackage{cancel}
\usepackage{slashed}
\usepackage{multirow}
\usepackage{slashed}
\usepackage{comment}
\usepackage{relsize}
\usepackage{tcolorbox}
\usepackage{csquotes}
\usepackage{simpler-wick}
\usepackage{lipsum}
\usepackage[compat=1.0.0]{tikz-feynman}
\usepackage{ulem}
\usepackage{subfigure}

\newcommand{\rmi}{{\rm i}}

\usepackage{ulem}

\allowdisplaybreaks

\begin{document}
\title{Complete Two-loop Renormalization Group Equation of the Weinberg Operator}

\author{Alejandro Ibarra}
\email{ibarra@tum.de}
\affiliation{
Technical University of Munich,
TUM School of Natural Sciences,
Department of Physics,
James-Franck-Stra\ss e,
85748 Garching,
Germany}

\author{Nicholas Leister}
\email{nleister@uni-mainz.de}
\affiliation{
Technical University of Munich,
TUM School of Natural Sciences,
Department of Physics,
James-Franck-Stra\ss e,
85748 Garching,
Germany}
\affiliation{
PRISMA+ Cluster of Excellence and Mainz Institute for Theoretical Physics,
Johannes Gutenberg-Universität Mainz, 
55099 Mainz, 
Germany}

\author{Di Zhang}
\email[\href{mailto:di1.zhang@tum.de}{di1.zhang@tum.de} \textcolor{black}{(corresponding author)}]{}
\affiliation{
Technical University of Munich,
TUM School of Natural Sciences,
Department of Physics,
James-Franck-Stra\ss e,
85748 Garching,
Germany}
\begin{abstract}
We calculate the renormalization group equation (RGE) of the lepton-number-violating Weinberg operator with the particle content of the Standard Model (SM), thus completing the set of two-loop RGEs of the SM effective field theory up to dimension 5. We identify new diagrams that could increase the rank of the Wilson coefficient of the Weinberg operator, and we calculate the complete two-loop RGE for the neutrino mass eigenvalues and leptonic mixing matrix. We also briefly discuss some phenomenological implications of the RGEs. 
\end{abstract}

\maketitle

\section{Introduction}

Neutrino oscillation experiments have established that the masses and mixing angles in the neutrino sector are qualitatively very different from those in the quark sector. A likely explanation for these differences is that neutrinos could be Majorana fermions rather than Dirac fermions. The lowest operator invariant under the Standard Model (SM) gauge group and containing only SM particles is the (dimension-5) Weinberg operator~\cite{Weinberg:1979sa}. In the simplest model realizations, the Weinberg operator is generated by integrating-out heavy degrees of freedom coupling to the left-handed lepton doublets (for early works, see Refs.~\cite{Minkowski:1977sc,Yanagida:1979as,Gell-Mann:1979vob,Mohapatra:1979ia,Schechter:1980gr,Foot:1988aq}). Since this operator is generated at a high energy scale, the overall size and the flavor structure of the matricial Wilson coefficient of the Weinberg operator can be affected by quantum effects. Given the large hierarchy between the scale at which the Weinberg operator is generated and the scale at which neutrino experiments are done, the leading quantum effects are encoded in the Renormalization Group Equations (RGEs).

The one-loop RGE of the Weinberg operator in the Standard Model was first calculated in Refs.~\cite{Chankowski:1993tx,Babu:1993qv,Antusch:2001ck} (for the RGE in extensions of the SM, see Refs.~\cite{Babu:1993qv,Ibarra:1999se,Antusch:2002ek,Grimus:2004yh}). The phenomenological implications of the RGEs have been discussed in various works, see {\it e.g.} Refs.~\cite{Casas:1999tp,Ellis:1999my,Casas:1999kc,Casas:1999tg,Chankowski:1999xc,Haba:1999ca,Haba:1999ca,Antusch:2001ck,Ohlsson:2013xva}. The RGEs can change the values of the neutrino mass eigenvalues and the mixing angles from the high energy cut-off scale where the Weinberg operator is plausibly generated to the low energies where the experiments are performed. Concretely, it has been pointed out that for degenerate neutrino mass eigenvalues at the cut-off scale, quantum effects could generate a splitting between these two mass eigenvalues and could lead to quasi-fixed points in the infrared for some elements of the leptonic mixing matrix, leading to a high predictivity for these scenarios. Furthermore, it has been shown that one-loop RGEs cannot generate a non-vanishing neutrino mass eigenvalue \cite{Mei:2003gn,Antusch:2005gp,Zhang:2020lsd,Zhang:2024weq}. On the other hand, while quark masses are protected by the chiral symmetry against large quantum effects, this is not the case of Majorana masses (when lepton flavor is violated). Therefore, one could expect the generation of a non-zero neutrino mass at a higher order in perturbation theory, even if this mass is equal to zero at a high energy scale. 

The two-loop RGE of the Weinberg operator in the Standard Model has been partially discussed in Ref.~\cite{Davidson:2006tg}, confirming that neutrino masses can be generated radiatively at the two-loop level (see also Refs.~\cite{Babu:1988ig,Choudhury:1994vr,Ma:1998db,Ibarra:2018dib,Ibarra:2020eia} for related analyses). The full two-loop RGE of the Weinberg operator has been calculated in Ref.~\cite{Antusch:2002ek} for the Minimal Supersymmetric Standard Model, exploiting the non-renormalization theorems in supersymmetry~\cite{Wess:1973kz,Iliopoulos:1974zv} that greatly simplify the calculation. On the other hand, these same non-renormalization theorems forbid the possibility of generating radiatively a non-zero neutrino mass, unless supersymmetry is broken~\cite{Davidson:2006tg}. In this work, we calculate for the first time the complete two-loop RGE of the Weinberg operator with the SM particle content, aiming to complete the program initiated in the 1980s of calculating the full RGEs of all couplings of the Standard Model~\cite{Machacek:1983fi,Machacek:1983tz,Machacek:1984zw,Luo:2002ey,Luo:2002ti}, to include also the simplest gauge invariant Lagrangian generating neutrino masses,  {\it i.e.} the Weinberg operator.

The remainder of this paper is organized as follows. In Section~\ref{sec:framework}, we briefly introduce the renormalization of the Weinberg operator using the Background Field Method (BFM) and we present the general formula to extract beta-functions from counterterms in the Dimension Regularization (DR) and the modified minimal subtraction ($\overline{\text{MS}}$) scheme. We first focus in  Section~\ref{sec:rank-increase}  on the contributions to the RGE that increase the rank of the neutrino mass matrix and then, in Section~\ref{sec:complete}, we derive the full two-loop RGE.  In Section~\ref{sec:mass-generation} we discuss some phenomenological implications of the two-loop RGEs and finally in Section~\ref{sec:conclusions} we present our conclusions.

\section{Renormalization of the Weinberg operator in the Background Field Method}\label{sec:framework}

The Lagrangian for the SM supplemented with the unique dimension-five Weinberg operator is~\cite{Weinberg:1979sa}
\begin{eqnarray}\label{eq:Ltot}
	\mathcal{L} = \mathcal{L}^{}_{\rm SM} + \left( \frac{1}{2} C^{\alpha\beta}_5 \overline{\ell^{}_{\alpha\rm L}} \widetilde{H} \widetilde{H}^{\rm T} \ell^{\rm c}_{\beta \rm L} + {\rm h.c.} \right)\;,
\end{eqnarray}
where $\ell^{}_{\rm L}$ are the lepton doublets, $\widetilde H= i\sigma_2 H^\ast_{}$ with $H$ being the Higgs doublet, and $\alpha,\beta=e,\mu,\tau$ are flavor indices. 

To simplify the renormalization of the gauge sector, we adopt the BFM~\cite{Abbott:1980hw,Abbott:1981ke,Abbott:1983zw}. In this framework, all fields are split into background and quantum parts, which respectively appear as external lines (or tree-level propagators) and loop propagators. Then the background-field effective action generating Green's functions becomes gauge invariant with respect to gauge transformations of background fields if the gauge fixing condition for quantum gauge fields is chosen as~\cite{Abbott:1980hw}
\begin{eqnarray}\label{eq:gc}
	\mathcal{G}^i_V = D^{}_\mu \hat{V}^{i\mu}= \partial_\mu\hat {V}^{i\mu} - \rmi g_V T^j_{ik} V^j_\mu \hat{V}^{k\mu} \;.
\end{eqnarray}
Here, $V$ ( $\hat{V}$) is any background (quantum) gauge field, and $g_V$ and $T^j$ are respectively the gauge coupling and the adjoint representation of the corresponding gauge group (with $j$ being the adjoint index). With this gauge fixing condition, one can readily write the gauge fixing and ghost field terms for the quantum gauge fields.
It should be noted that the gauge condition in Eq.~\eqref{eq:gc} does not contain any fermion and Higgs fields. Hence, all interactions of the background and quantum fields for fermions and Higgs are exactly the same and indistinguishable, making it unnecessary to split the fermions and Higgs fields into a background and quantum component. Consequently, we only need to split gauge fields, including those in covariant derivatives in Eq.~\eqref{eq:Ltot} into background and quantum parts, which will not be shown explicitly here. More details can be found in Refs.~\cite{Abbott:1980hw,Abbott:1981ke,Abbott:1983zw}.

In the BFM, the quantum fields, as well as ghost fields, always appear in loops, and hence wave-function renormalization constants associated with propagators and the adjacent vertices are canceled out. As a result, the renormalization of quantum and ghost fields is unnecessary. Moreover, the renormalization constants for the background gauge fields and the corresponding gauge couplings are related by the gauge invariance, namely~\cite{Abbott:1980hw}
\begin{eqnarray}
	Z^{}_{g^{}_1} = Z^{-1/2}_B \;,\quad Z^{}_{g^{}_2} = Z^{-1/2}_W \;,\quad Z^{}_{g^{}_3} = Z^{-1/2}_G \;.
\end{eqnarray}
Accordingly, the wave-function renormalization of background gauge fields is sufficient to renormalize the whole gauge sector, which is much simpler than the conventional renormalization of a gauge theory. However, the gauge-fixing parameters for quantum gauge fields still need to be renormalized. Hence, besides the renormalization constants for the SM couplings and background fields, one has to introduce those in $d = 4 -2\varepsilon$ dimensional space-time for the gauge-fixing parameters and the Wilson coefficient of the Weinberg operator,~{\it i.e.,}
\begin{eqnarray}\label{eq:renorm-cont}
    &&\xi^{}_{\hat{V}} = Z^{}_{\xi_{\hat{V}}} \xi^{}_{\hat{V},r} \;,\quad C^{\alpha\beta}_5 = \mu^{2\varepsilon} \left( C^{\alpha\beta}_{5,r} + \delta C^{\alpha\beta}_5 \right) \;,
\end{eqnarray}
with $\mu$ and the subscript $r$ denoting an arbitrary parameter of mass-dimension one and the renormalized quantity, respectively. Then the bare Lagrangian can be written into a renormalized Lagrangian with all counterterms. The explicit expressions of those counterterms or renormalization constants can be obtained by calculating relevant 1PI Green's functions and requiring the cancellation of all divergences at each loop level. Due to the independence of the bare quantities to the renormalization scale $\mu$, one can derive renormalization group equations for all couplings from the corresponding renormalization constants.

In the $d = 4 -2\varepsilon$ DR and the $\overline{\text{MS}}$ scheme, the renormalized couplings $\kappa^{}_{i,r}$ are related to the corresponding bare couplings $\kappa^{}_{i}$ by an expansion of the form
\begin{eqnarray}\label{eq:expansion}
	\kappa^{}_i \mu^{-\rho^{}_{k_i} \varepsilon} &=& \kappa^{}_{i,r} + \sum^\infty_{n=1} a^{(n)}_i (\kappa^{}_r) \frac{1}{\varepsilon^n}
\end{eqnarray}  
where the expansion coefficients $a^{(n)}_i$ can be obtained from the renormalization constants $\delta Z^{}_{\kappa_i}$ of $\kappa^{}_i$. From Eq.~\eqref{eq:expansion}, one can obtain the beta function of $\kappa^{}_{i,r}$, namely,
\begin{eqnarray}\label{eq:beta-function}
	\mu\frac{{\rm d}\kappa^{}_{i,r}}{{\rm d}\mu} &=& \beta^{}_i - \rho^{}_{\kappa_i} \kappa^{}_{i,r} \varepsilon 
\end{eqnarray}
and a relation between expansion coefficients, {\it i.e.,}
\begin{eqnarray}\label{eq:iterate-relation}
	a^{(n+1)}_i = - \frac{1}{\rho^{}_{\kappa_i}} \sum^{}_j \left( \beta^{}_j \frac{\partial a^{(n)}_i }{\partial \kappa^{}_{j,r}} - \rho^{}_{\kappa_j} \kappa^{}_{j,r} \frac{\partial a^{(n+1)}_i}{\partial \kappa^{}_{j,r}} \right)
\end{eqnarray}
with 
\begin{eqnarray}\label{eq:beta}
	\beta^{}_i = \sum^{}_j \rho^{}_{\kappa_j} \kappa^{}_{j,r} \frac{\partial a^{(1)}_i}{\partial \kappa^{}_{j,r}} - \rho^{}_{\kappa^{}_i} a^{(1)}_i \;.
\end{eqnarray}
As can be seen from Eqs.~\eqref{eq:expansion} and \eqref{eq:beta}, the beta functions are fully determined by the expansion coefficient of the first pole of $\varepsilon$. Meanwhile, Eq.~\eqref{eq:iterate-relation} indicates that expansion coefficients of the higher-order poles are completely determined by those of the lower-order poles and provides a useful computational check. Eqs.~\eqref{eq:expansion}-\eqref{eq:beta} can be easily generalized to coupling matrices with flavor indices. In this work, we will employ the Feyman-'t Hooft gauge $\xi^{}_{\hat{V},r} = 1$ and drop henceforth the subscript $r$ of all renormalized quantities for ease in the notation.

\section{Rank-increasing Contributions}\label{sec:rank-increase}

\begin{figure}[t!]
	\centering
	\includegraphics[width=\linewidth]{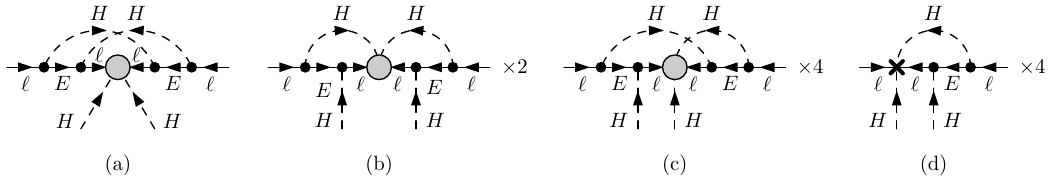}
	\vspace{-0.6cm}
	\caption{Diagrams increasing the rank of neutrino mass matrix at the two-loop level. The last diagram results from the one-loop counterterms.}
	\label{fig:fd}
\end{figure}

The two-loop RGE of the Weinberg operator has been discussed in Ref.~\cite{Davidson:2006tg} (see also Ref.~\cite{Xing:2020ezi} for more detailed numerical analysis), focusing on the possibility of generating radiatively a non-zero neutrino mass via contributions of the form $( Y^{}_l Y^\dagger_l ) C^{}_5 ( Y^{}_l Y^\dagger_l )^{\rm T} $. We show in Fig.~\ref{fig:fd} all possible topologies leading to RGE terms with that structure, of which only Diagram (a) was considered in Ref.~\cite{Davidson:2006tg}. In this section, we will revisit the calculation of the rank-increasing contributions to the RGE, including all the diagrams in Fig.~\ref{fig:fd}.~\footnote{Two-loop contributions from the exchange of two $W$ bosons~\cite{Petcov:1984nz,Babu:1988ig,Choudhury:1994vr,Ma:1998db} could also increase the rank of the mass matrix in the presence of sterile neutrinos. 
}

Since for the calculation of the RGEs we are interested only in the divergent part of the diagram, one can take all external momenta to zero when calculating the diagrams in Fig.~\ref{fig:fd}. For example, the explicit expression for the amplitude of Diagram (a) is:
\begin{eqnarray}\label{eq:amp1}
	\mathcal{M}_a &=& \rmi \left( \epsilon^{ac}\epsilon^{bd}+\epsilon^{ad}\epsilon^{bc} \right) \left( Y^{}_l Y^\dagger_l \right)^{}_{\rho\beta} \left( C^{\dagger}_5 \right)^{\rho\lambda} \left( Y^{}_l Y^\dagger_l \right)^{}_{\lambda\alpha}  
	\nonumber
	\\
	&&\times \overline{u}(0) P^{}_{\rm L} \int \frac{{\rm d}^d k^{}_1}{\left( 2\pi \right)^d} \int \frac{{\rm d}^d k^{}_2}{\left( 2\pi \right)^d} \frac{ \slashed{k}^{}_2 \slashed{k}^{}_1 } {k^2_1 k^2_2  \left( k^{}_2 - k^{}_1 \right)^2 \left( k^2_1 -M^2 \right) \left( k^2_2 - M^2 \right)} u(0)
	\nonumber
	\\
	&=&  \rmi \left( \epsilon^{ac}\epsilon^{bd}+\epsilon^{ad}\epsilon^{bc} \right) \left( Y^{}_l Y^\dagger_l \right)^{}_{\rho\beta} \left( C^{\dagger}_5 \right)^{\rho\lambda} \left( Y^{}_l Y^\dagger_l \right)^{}_{\lambda\alpha} \overline{u}(0) P^{}_{\rm L} u(0) 
	\nonumber
	\\
	&& \times \frac{1}{\left( 4\pi \right)^d} \left[ \frac{1}{2\left( d- 3\right) M^4} \left( A^{(d)}_{\{1,M\}} \right)^2 - \frac{1}{M^2} K^{(d)}_{\{1,M\},\{1,0\},\{1,0\}} \right]
	\nonumber
	\\
	&\supset&  \rmi \left( \epsilon^{ac}\epsilon^{bd}+\epsilon^{ad}\epsilon^{bc} \right) \left( Y^{}_l Y^\dagger_l \right)^{}_{\rho\beta} \left( C^{\dagger}_5 \right)^{\rho\lambda} \left( Y^{}_l Y^\dagger_l \right)^{}_{\lambda\alpha} \overline{u}(0) P^{}_{\rm L} u(0) \times \left[ - \frac{1}{\left( 16\pi^2 \right)^2} \frac{1}{2\varepsilon} \right] \;,
\end{eqnarray}
where we have used the Feynman rules for fermion-number-violating interactions in Ref.~\cite{Denner:1992vza}, while the definitions of loop integrals $A^{(d)}_{\{1,M\}}$ and $K^{(d)}_{\{1,M\},\{1,0\},\{1,0\}}$ are included in Appendix~\ref{app:vertex} together with their explicit results. Similarly, one can calculate the remaining diagrams in Fig.~\ref{fig:fd}. We find that the two-loop vertex renormalization constant $\delta V^{(2)}_5$ proportional  to $( Y^{}_l Y^\dagger_l ) C^{}_5 ( Y^{}_l Y^\dagger_l )^{\rm T} $ is:
\begin{eqnarray}\label{eq:two-loop-ct}
	\delta V^{(2)}_5 &=& \delta V^{(2)}_{5,a} + \delta V^{(2)}_{5,b} + \delta V^{(2)}_{5,c} + \delta V^{(2)}_{5,d} + {\dots} = \frac{1}{\left( 16\pi^2 \right)^2} \left( \frac{1}{\varepsilon^2} + \frac{1}{2\varepsilon} \right) \left( Y^{}_l Y^\dagger_l \right) C^{}_5 \left( Y^{}_l Y^\dagger_l \right)^{\rm T} + \dots \quad\;
\end{eqnarray}
where the contributions from Diagrams (a)-(d) are:
\begin{eqnarray}\label{eq:v5-res}
	\delta V^{(2)}_{5,a} &=& \frac{1}{\left( 16\pi^2 \right)^2} \frac{1}{2\varepsilon} \left( Y^{}_l Y^\dagger_l \right) C^{}_5 \left( Y^{}_l Y^\dagger_l \right)^{\rm T} \;,
	\nonumber
	\\
	\delta V^{(2)}_{5,b} &=& - \frac{1}{\left( 16\pi^2 \right)^2} \frac{2}{\varepsilon^2} \left( 1 + 2\varepsilon  - 2 \varepsilon\Delta  \right)  \left( Y^{}_l Y^\dagger_l \right) C^{}_5 \left( Y^{}_l Y^\dagger_l \right)^{\rm T} \;,
	\nonumber
	\\
	\delta V^{(2)}_{5,c} &=&  \frac{1}{\left( 16\pi^2 \right)^2} \frac{1}{\varepsilon^2} \left( 1 + 2\varepsilon  - 2 \varepsilon\Delta \right)  \left( Y^{}_l Y^\dagger_l \right) C^{}_5 \left( Y^{}_l Y^\dagger_l \right)^{\rm T} \;,
	\nonumber
	\\
	\delta V^{(2)}_{5,d} &=& \frac{1}{\left( 16\pi^2 \right)^2} \frac{2}{\varepsilon^2} \left( 1 + \varepsilon  -  \varepsilon\Delta  \right)  \left( Y^{}_l Y^\dagger_l \right) C^{}_5 \left( Y^{}_l Y^\dagger_l \right)^{\rm T} \;,
\end{eqnarray}
with $\Delta = \gamma^{}_{\rm E} - \ln(4\pi) + \ln(M^2/\mu^2)$. As expected, the logarithmic terms $\ln (M^2/\mu^2)$ cancel out when summing all diagrams. We then calculate the two-loop counterterm $\delta C^{(2)}_5$ by subtracting contributions of wave-function renormalization constants from $\delta V^{(2)}_5$, {\it i.e.,}
\begin{eqnarray}\label{eq:c5-res}
	\delta C^{(2)}_5 &=& \delta V^{(2)}_5 - \frac{1}{2} \delta Z^{(1)}_\ell \delta C^{(1)}_5  - \frac{1}{2} \delta C^{(1)}_5 \left( \delta Z^{(1)}_\ell \right)^{\rm T} - \frac{1}{4} \delta Z^{(1)}_\ell C^{}_5 \left( \delta Z^{(1)}_\ell \right)^{\rm T} + \dots
	\nonumber
	\\
	&=& \frac{1}{\left( 16\pi^2 \right)^2} \left( \frac{9}{16\varepsilon^2} + \frac{1}{2\varepsilon} \right) \left( Y^{}_l Y^\dagger_l \right) C^{}_5 \left( Y^{}_l Y^\dagger_l \right)^{\rm T} + \dots
\end{eqnarray}
where the one-loop counterterms are listed in Appendix~\ref{app:one-loop}. Note that only Diagram (a) contributes to the $1/\varepsilon$ part of the counterterm, and that the sum of the Diagrams (b)-(d) as well as the product of one-loop counterterms only gives contributions proportional to $1/\varepsilon^2$. As a consequence, only Diagram~(a) will contribute to the two-loop RGE in accordance with the results of Ref.~\cite{Davidson:2006tg}. However, the inclusion of the extra diagrams and of the one-loop counterterms are crucial as a consistency check of the result described in Eq.~\eqref{eq:iterate-relation}.
Concretely, the expansion factors $a^{(1)}$ and $a^{(2)}$ for $C^{}_5$ are 
\begin{eqnarray}\label{eq:cf-pole}
	a^{(1)} &=& - \frac{1}{16\pi^2}  \frac{3}{4} \left[ \left( Y^{}_l Y^\dagger_l \right) C^{}_5 + C^{}_5 \left( Y^{}_l Y^\dagger_l \right)^{\rm T} \right] + \dots \;,\nonumber \\
	a^{(2)} &=&  \frac{1}{\left( 16\pi^2 \right)^2} \frac{9}{16}  \left( Y^{}_l Y^\dagger_l \right) C^{}_5 \left( Y^{}_l Y^\dagger_l \right)^{\rm T} + \dots \;,\quad
\end{eqnarray}
where the two-loop contribution in $a^{(1)}$ has been omitted. Replacing them in Eq.~\eqref{eq:iterate-relation} one finds that the expected relation 
\begin{eqnarray}\label{eq:ing}
	a^{(2)}=-\frac{1}{2} \sum^{}_j  \left( \beta^{}_j \frac{\partial a^{(1)} }{\partial \kappa^{}_j} - \rho^{}_{\kappa_j} \kappa^{}_j \frac{\partial a^{(2)}}{\partial \kappa^{}_j}  \right) = -\frac{1}{2}  \frac{1}{\left( 16\pi^2 \right)^2} \left( \frac{9}{4} - \frac{27}{8} \right) \left( Y^{}_l Y^\dagger_l \right) C^{}_5 \left( Y^{}_l Y^\dagger_l \right)^{\rm T}  + \dots \;,
\end{eqnarray}
is indeed satisfied. 

\section{Complete Two-loop RGE of the Weinberg Operator}\label{sec:complete}

In this section we calculate for the first time the complete RGE of the Weinberg operator up to two-loops. To this end, we calculate the divergent parts of all diagrams renormalizing $C^{}_5$ up to two-loops taking all external momenta to zero, reducing them to vacuum diagrams.~\footnote{We used {\sf FeynRules}~\cite{Christensen:2008py,Alloul:2013bka} to generate the Feynman rules, {\sf FeynArts}~\cite{Hahn:2000kx} to generate the Feynman diagrams, {\sf FeynCalc}~\cite{Shtabovenko:2016sxi,Shtabovenko:2020gxv} to calculate the amplitudes, and {\sf TARCER}~\cite{Mertig:1998vk} to reduce the tensor integrals.}
We determine not only the  $1/\varepsilon$ pole of the diagrams, but also the  $1/\varepsilon^2$ pole, in order to apply Eq.~\eqref{eq:iterate-relation} as a check of our result. 

As is well known, the reduction of the diagrams generates spurious infrared divergences entangled with the ultraviolet divergences. We follow the approach proposed in Ref.~\cite{Chetyrkin:1997fm} to rearrange the infrared divergences by introducing a unique artificial mass for some relevant fields, that we denote by $M$. Since the quadratic term for the Higgs doublet in the Lagrangian is not relevant for our calculations, we can regard it as a spurious mass term to regularize infrared divergences. Besides, we also introduce the same spurious mass $M$ for both the quantum $\hat{B}$ and $\hat{W}$ gauge fields (see also Refs.~\cite{Zoller:2014xoa,Lang:2020nnl,Brod:2020lhd,Jenkins:2023rtg} for recent implementations of this method). Note that even though they have a common mass $M$, their mass renormalization constants could be different due to their different interactions. In our calculation, introducing spurious mass terms for $\hat{B}$, $\hat{W}$ and the Higgs doublet suffices to rearrange all infrared divergences~\cite{Chetyrkin:1997fm}.

To determine the one-loop RGE of the Weinberg operator, we calculate the one-loop counterterm of the Wilson coefficient
\begin{eqnarray}\label{eq:one-loop-C5}
\delta C^{(1)}_5 &=& \delta V^{(1)}_5 - \frac{1}{2}  \delta Z^{(1)}_\ell C^{}_5 - \frac{1}{2} C^{}_5 \delta Z^{(1)\rm T}_\ell - \delta Z^{(1)}_H C^{}_5\,
\end{eqnarray}
where the one-loop Weinberg operator vertex counterterm is
\begin{eqnarray}
    \delta V^{(1)}_5 = \frac{1}{16\pi^2 \varepsilon }  \left[ \left( 2\lambda + \frac{1}{4} g^2_1 -\frac{3}{4} g^2_2 \right)C^{}_5  -  Y^{}_l Y^\dagger_l C^{}_5 -  C^{}_5 \left( Y^{}_l Y^\dagger_l \right)^{\rm T} \right]  \;,
\end{eqnarray}
and the wave-function renormalization constants of lepton and Higgs doublets are given in Eq.~\eqref{eq:one-loop-results}. Replacing in Eq.~(\ref{eq:one-loop-C5}), we find
\begin{eqnarray}\label{eq:one-loop-C5-ct}
\delta C^{(1)}_5 
&=& \frac{1}{16\pi^2 \varepsilon \cdot 4}  \left[ 2\left( 4\lambda -3g^2_2 + 2T \right)C^{}_5  - 3 Y^{}_l Y^\dagger_l C^{}_5 - 3 C^{}_5 \left( Y^{}_l Y^\dagger_l \right)^{\rm T} \right] \;,
\end{eqnarray}
with $T = {\rm Tr} \left( Y^{}_l Y^\dagger_l + 3Y^{}_{\rm u} Y^\dagger_{\rm u} + 3Y^{}_{\rm d}Y^\dagger_{\rm d}  \right)$.

Similarly, to determine the two-loop RGE of the Weinberg operator we first calculate the two-loop counterterm for the Wilson coefficient $C_5$:
\begin{eqnarray}
	\delta C^{(2)}_5 &=& \delta V^{(2)}_5 - \frac{1}{2} \delta Z^{(2)}_\ell C^{}_5 - \frac{1}{2} C^{}_5 \left( \delta Z^{(2)}_\ell \right)^{\rm T} - \delta Z^{(2)}_H C^{}_5 - \frac{1}{2} \delta Z^{(1)}_\ell \delta C^{(1)}_5  - \frac{1}{2} \delta C^{(1)}_5 \left( \delta Z^{(1)}_\ell \right)^{\rm T} + \frac{1}{8}  \delta Z^{(1)}_\ell  \delta Z^{(1)}_\ell C^{}_5 
	\nonumber
	\\
	&& + \frac{1}{8} C^{}_5 \left(  \delta Z^{(1)}_\ell  \delta Z^{(1)}_\ell \right)^{\rm T} - \frac{1}{2} \delta Z^{(1)}_H \left[ 2\delta C^{(1)}_5 + \delta Z^{(1)}_\ell C^{}_5 + C^{}_5  \left(\delta Z^{(1)}_\ell \right)^{\rm T} \right] - \frac{1}{4} \delta Z^{(1)}_\ell C^{}_5 \left( \delta Z^{(1)}_\ell \right)^{\rm T} \;,
\end{eqnarray}
which depends on the one-loop counterterm $\delta C_5^{(1)}$, the two-loop Weinberg operator vertex counterterms, $\delta V_5^{(2)}$, and the one and two-loop wave-function renormalization constants of the lepton and Higgs doublets, $\delta Z_\ell^{(1)}$, $\delta Z_H^{(1)}$, $\delta Z_\ell^{(2)}$, $\delta Z_H^{(2)}$. The details of the calculation and their explicit expressions are given in Appendices~\ref{app:one-loop}, \ref{app:vertex} and \ref{app:lepton-Higgs}.
Substituting all counterterms in the above equation one obtains
\begin{eqnarray}\label{eq:two-loop-ct-c5}
	\delta C^{(2)}_5 &=& \frac{1}{\left( 16\pi^2 \right)^2  \varepsilon^2} \left\{ \left[\frac{3}{16} g^4_1 + \frac{3}{8} g^2_1 g^2_2 + \frac{65}{16} g^4_2 - \lambda \left( \frac{3}{2} g^2_1 + \frac{15}{2} g^2_2 - 14\lambda - 4T \right) - \frac{1}{8} g^2_1 {\rm Tr} \left( 15Y^{}_l Y^\dagger_l  \right. \right.\right.
	\nonumber
	\\
	&& + \left.\left. 17 Y^{}_{\rm u} Y^\dagger_{\rm u} + 5Y^{}_{\rm d} Y^\dagger_{\rm d} \right)  - \frac{21}{8} g^2_2 T - 12g^2_3 {\rm Tr} \left( Y^{}_{\rm u} Y^\dagger_{\rm u} + Y^{}_{\rm d} Y^\dagger_{\rm d}\right) + T^2 - \frac{1}{4} T^\prime - \frac{9}{2} {\rm Tr} \left( Y^{}_{\rm u} Y^\dagger_{\rm u} Y^{}_{\rm d} Y^\dagger_{\rm d} \right) \right] C^{}_5
	\nonumber
	\\
	&& + \left( \frac{45}{32} g^2_1 + \frac{63}{32} g^2_2 - \frac{3}{2} \lambda - \frac{9}{8} T \right) \left[ Y^{}_l Y^\dagger_l C^{}_5 + C^{}_5 \left( Y^{}_l Y^\dagger_l \right)^{\rm T} \right] - \frac{9}{32} \left[Y^{}_l Y^\dagger_l Y^{}_l Y^\dagger_l C^{}_5 + C^{}_5 \left( Y^{}_l Y^\dagger_l Y^{}_l Y^\dagger_l \right)^{\rm T} \right]
	\nonumber
	\\
	&& + \left. \frac{9}{16} Y^{}_l Y^\dagger_l C^{}_5 \left( Y^{}_l Y^\dagger_l \right)^{\rm T} \right\}
	\nonumber
	\\
	&& + \frac{1}{\left( 16\pi^2 \right)^2  \varepsilon} \left\{ - \left[ \frac{129}{32} g^4_1 + \frac{83}{16} g^2_1 g^2_2 +  \frac{169}{96} g^4_2 + \lambda \left( g^2_1 + 7\lambda + 2 T \right) - \frac{15}{16} g^2_2 T - 10g^2_3 {\rm Tr} \left( Y^{}_{\rm u} Y^\dagger_{\rm u} + Y^{}_{\rm d} Y^\dagger_{\rm d}\right) \right.\right.
	\nonumber
	\\
	&& - \left. \frac{5}{48} g^2_1 {\rm Tr} \left( 15 Y^{}_l Y^\dagger_l + 17 Y^{}_{\rm u} Y^\dagger_{\rm u} + 5 Y^{}_{\rm d} Y^\dagger_{\rm d} \right) + \frac{1}{8} T^\prime - \frac{3}{4} {\rm Tr} \left( Y^{}_{\rm u} Y^\dagger_{\rm u} Y^{}_{\rm d} Y^\dagger_{\rm d} \right) \right] C^{}_5 - \left( \frac{57}{64} g^2_1 -  \frac{33}{64} g^2_2 - \frac{5}{16} T \right) 
	\nonumber
	\\
	&& \times \left[ Y^{}_l Y^\dagger_l C^{}_5 + C^{}_5 \left( Y^{}_l Y^\dagger_l \right)^{\rm T} \right] + \frac{19}{16} \left[Y^{}_l Y^\dagger_l Y^{}_l Y^\dagger_l C^{}_5 + C^{}_5 \left( Y^{}_l Y^\dagger_l Y^{}_l Y^\dagger_l \right)^{\rm T} \right] + \left. \frac{1}{2} Y^{}_l Y^\dagger_l C^{}_5 \left( Y^{}_l Y^\dagger_l \right)^{\rm T} \right\} \;,
\end{eqnarray}
where $T^\prime = {\rm Tr} \left[ \left( Y^{}_l Y^\dagger_l \right)^2 + 3\left( Y^{}_{\rm u} Y^\dagger_{\rm u} \right)^2 + 3 \left( Y^{}_{\rm d} Y^\dagger_{\rm d} \right)^2 \right]$. Finally, from the 1/$\varepsilon$ terms in $\delta C^{(1)}_5$ and $\delta C^{(2)}_5$, and using Eq.~\eqref{eq:beta}, we obtain the one- and two-loop contributions to the RGE of $C^{}_5$:
\begin{eqnarray}\label{eq:one-loop-rge}
	\mu \frac{{\rm d} C^{(1)}_5}{{\rm d} \mu} &=&  \frac{1}{16\pi^2}  \left[ \left( 4\lambda -3g^2_2 + 2T \right) C^{}_5 - \frac{3}{2} Y^{}_l Y^\dagger_l C^{}_5 - \frac{3}{2} C^{}_5 \left( Y^{}_l Y^\dagger_l \right)^{\rm T} \right] \;,
	\\ \label{eq:two-loop-rge}
	\mu \frac{{\rm d} C^{(2)}_5}{{\rm d} \mu} &=& \frac{1}{\left( 16\pi^2 \right)^2} \left\{ - \left[ \frac{129}{8}  g^4_1 + \frac{83}{4} g^2_1 g^2_2 + \frac{169}{24} g^4_2 + 4\lambda \left( g^2_1  + 7\lambda + 2T \right)   - \frac{15}{4} g^2_2 T - 40g^2_3 {\rm Tr} \left( Y^{}_{\rm u} Y^\dagger_{\rm u} + Y^{}_{\rm d} Y^\dagger_{\rm d}\right)  \right.\right.
	\nonumber
	\\
	&& - \frac{5}{12} g^2_1 {\rm Tr} \left( 15Y^{}_l Y^\dagger_l + 17 Y^{}_{\rm u} Y^\dagger_{\rm u} + 5 Y^{}_{\rm d} Y^\dagger_{\rm d} \right) + \left. \frac{1}{2} T^\prime -3 {\rm Tr} \left( Y^{}_{\rm u} Y^\dagger_{\rm u} Y^{}_{\rm d} Y^\dagger_{\rm d} \right) \right] C^{}_5 - \left( \frac{57}{16} g^2_1 -  \frac{33}{16} g^2_2 - \frac{5}{4} T \right) 
	\nonumber
	\\
	&& \times \left[ Y^{}_l Y^\dagger_l C^{}_5 + C^{}_5 \left( Y^{}_l Y^\dagger_l \right)^{\rm T} \right] + \frac{19}{4} \left[Y^{}_l Y^\dagger_l Y^{}_l Y^\dagger_l C^{}_5 + C^{}_5 \left( Y^{}_l Y^\dagger_l Y^{}_l Y^\dagger_l \right)^{\rm T} \right] + \left. 2 Y^{}_l Y^\dagger_l C^{}_5 \left( Y^{}_l Y^\dagger_l \right)^{\rm T} \right\} \;.
\end{eqnarray}
Our one-loop result is consistent with Refs.~\cite{Chankowski:1993tx,Babu:1993qv} including the correction pointed out in Ref.~\cite{Antusch:2001ck}. Together with the two-loop RGEs for the SM couplings~\cite{Machacek:1983tz,Machacek:1983fi,Machacek:1984zw,Luo:2002ey,Luo:2002ti} collected in Appendix~\ref{app:sm}, the full two-loop RGEs of the SM extended with the Weinberg operator ({\it i.e.,} the SMEFT up to dimension five) are now complete.

\section{The Smallest Neutrino Mass and the Associated Majorana Phase}\label{sec:mass-generation}
After spontaneous symmetry breaking the Weinberg operator generates a Majorana mass term for neutrinos, given by:
\begin{eqnarray}
	\mathcal{L}^{}_{\rm mass} = - \frac{1}{2} \overline{\nu^{}_{\rm L}} M^{}_\nu \nu^{\rm c}_{\rm L}
\end{eqnarray}
where $M^{}_\nu = - v^2 C^{}_5/2$ and $v$ is the Higgs vacuum expectation value. Using Eqs.~\eqref{eq:one-loop-rge} and \eqref{eq:two-loop-rge}, one obtains the RGE of the neutrino mass matrix up to the two-loop level, which can be cast as:
\begin{eqnarray}\label{eq:mass-matrix-rge}
	\frac{{\rm d} M^{}_\nu }{{\rm d} t} = \alpha M^{}_\nu + Q M^{}_\nu + M^{}_\nu Q^{\rm T} + 2 P M^{}_\nu P^{\rm T} \;,
\end{eqnarray}
with $t=\ln (\mu/\Lambda)$ and  $\Lambda$ the cut-off of the theory, while
\begin{eqnarray}
	\alpha &=& \frac{1}{16\pi^2} \left( 4\lambda -3g^2_2 + 2T \right) - \frac{1}{\left( 16\pi^2 \right)^2}  \left[ \frac{129}{8}  g^4_1 + \frac{83}{4} g^2_1 g^2_2 + \frac{169}{24} g^4_2 + 4\lambda \left( g^2_1  + 7\lambda + 2T \right)   - \frac{15}{4} g^2_2 T  + \frac{1}{2} T^\prime  \right.
	\nonumber
	\\
	&& - \left. \frac{5}{12} g^2_1 {\rm Tr} \left( 15Y^{}_l Y^\dagger_l + 17 Y^{}_{\rm u} Y^\dagger_{\rm u} + 5 Y^{}_{\rm d} Y^\dagger_{\rm d} \right) - 40g^2_3 {\rm Tr} \left( Y^{}_{\rm u} Y^\dagger_{\rm u} + Y^{}_{\rm d} Y^\dagger_{\rm d}\right)  -3 {\rm Tr} \left( Y^{}_{\rm u} Y^\dagger_{\rm u} Y^{}_{\rm d} Y^\dagger_{\rm d} \right) \right] \;,
	\nonumber
	\\
	P &=& \frac{1}{16\pi^2} Y^{}_l Y^\dagger_l \;,
	\nonumber
	\\
	Q &=& - \left[ \frac{3}{2} + \frac{1}{16\pi^2} \left( \frac{57}{16} g^2_1 -  \frac{33}{16} g^2_2 - \frac{5}{4} T \right) \right] P + \frac{19}{4} P^2 \;.
 \label{eq:definitions}
\end{eqnarray}

Working in the basis where the charged-lepton Yukawa coupling matrix $Y^{}_l$ is diagonal~\footnote{$Y^{}_l$ remains diagonal during the SM two-loop RG running if it is diagonal at the beginning. However, if there are higher dimensional operators, it may not keep diagonal even at the one-loop level due to the potential contributions from higher dimensional operators, see Refs.~\cite{Jenkins:2013zja,Wang:2023bdw} for an example.}, one can diagonalize the neutrino mass matrix via $U^\dagger M^{}_\nu U^\ast = D^{}_\nu \equiv {\rm diag} \left( m^{}_1, m^{}_2, m^{}_3 \right) $ with $m^{}_i$ (for $i=1,2,3$) being real and positive, and the unitary matrix $U$ is the leptonic mixing matrix, {\it i.e.}, the Pontecorvo-Maki-Nakagawa-Sakata (PMNS) matrix~\cite{Pontecorvo:1957cp,Maki:1962mu}. Replacing this decomposition of the neutrino mass matrix into Eq.~\eqref{eq:mass-matrix-rge} it follows that:
\begin{eqnarray}\label{eq:mass-matrix-rge-diag}
	\frac{{\rm d} D^{}_\nu }{{\rm d} t} = \alpha D^{}_\nu + \widehat{Q} D^{}_\nu + D^{}_\nu \widehat{Q}^{\rm T} + 2 \widehat{P} D^{}_\nu \widehat{P}^{\rm T} - \mathcal{T} D^{}_\nu + D^{}_\nu \mathcal{T}^\ast \;,
\end{eqnarray}
with $\widehat{Q} = U^\dagger Q U $, $\widehat{P} = U^\dagger P U $ and $\mathcal{T} = U^\dagger {\rm d} U/{\rm d} t $. Note that both $\widehat{Q}$ and $\widehat{P}$ are Hermitian matrices and $\mathcal{T}$ is an anti-Hermitian matrix, hence ${\rm Im} \left( \widehat{Q}^{}_{ii} \right), {\rm Im} \left( \widehat{P}^{}_{ii} \right), {\rm Re} \left( \mathcal{T}^{}_{ii} \right) =0$. Then, from the diagonal and non-diagonal elements of Eq.~\eqref{eq:mass-matrix-rge-diag} one obtains, respectively,
\begin{eqnarray}\label{eq:diagonal-rge1}
	\frac{{\rm d} m^{}_i }{{\rm d} t} &=& \left( \alpha + 2\widehat{Q}_{ii} \right) m^{}_i + 2 \sum^{}_k m^{}_k {\rm Re} \left(  \widehat{P}^2_{ik} \right) \;,
	\\\label{eq:diagonal-rge2}
	m^{}_i {\rm Im} \left( \mathcal{T}^{}_{ii} \right) &=& \sum^{}_k m^{}_k {\rm Im} \left( \widehat{P}^2_{ik} \right) \;,
\end{eqnarray}
and
\begin{eqnarray}\label{eq:nondiagonal-rge1}
	\left( m^{}_j - m^{}_i \right) {\rm Re} \left( \mathcal{T}^{}_{ij} \right) &=& \left( m^{}_j + m^{}_i \right)  {\rm Re} \left( \widehat{Q}^{}_{ij} \right) + 2 \sum^{}_k m^{}_k {\rm Re} \left( \widehat{P}^{}_{ik} \widehat{P}^{}_{jk} \right) \;,
	\\\label{eq:nondiagonal-rge2}
	\left( m^{}_j + m^{}_i \right) {\rm Im} \left( \mathcal{T}^{}_{ij} \right) &=& \left( m^{}_j - m^{}_i \right)  {\rm Im} \left( \widehat{Q}^{}_{ij} \right) + 2 \sum^{}_k m^{}_k {\rm Im} \left( \widehat{P}^{}_{ik} \widehat{P}^{}_{jk} \right) \;,
\end{eqnarray}
with $\widehat{P}^2_{ik} = ( \widehat{P}^{}_{ik})^2$.  Eq.~\eqref{eq:diagonal-rge1} determines the RG running of the neutrino mass eigenvalues, and Eqs.~\eqref{eq:diagonal-rge2}-\eqref{eq:nondiagonal-rge2} govern those of the PMNS matrix (the structure of the RGEs for neutrino masses and the PMNS matrix is analogous to the one for right-handed neutrino masses and the right-handed mixing matrix~\cite{Ibarra:2020eia}).

To investigate the impact of the RGEs on the mass eigenvalues and mass differences, let us first consider a simplified scenario with only two neutrinos, and later on the realistic case with three neutrinos.

\subsection{Two neutrino case}
A system of differential linear equations with non-constant coefficients does not admit in general a closed form. However, an analytic solution exists for constant coefficients. Neglecting the running of the coupling constants in Eq.(\ref{eq:definitions}), Eq.~\eqref{eq:diagonal-rge1} can be analytically solved, giving
\begin{align}
m_1(t_{\rm EW})&\simeq e^{(\xi_{11}+\xi_{22})/2}\Big[\cosh(\Omega) m_1(0)+ \frac{\sinh(\Omega)}{\Omega}\Big(\xi_{12} m_2(0) +\frac{\xi_{11}-\xi_{22}}{2}m_1(0)\Big)\Big], \nonumber \\
m_2(t_{\rm EW})&\simeq e^{(\xi_{11}+\xi_{22})/2}\Big[\cosh(\Omega) m_2(0)+ \frac{\sinh(\Omega)}{\Omega}\Big(\xi_{12} m_1(0)-\frac{\xi_{11}-\xi_{22}}{2}m_2(0) \Big)\Big],
\end{align}
where
\begin{align}
\xi_{ii}&\simeq \Big[\alpha+2\widehat Q_{ii}+2 \, {\rm Re}\Big(\widehat P^2_{ii}\Big)\Big] t_{\rm EW}, \nonumber\\
\xi_{12}&\simeq  2 \,{\rm Re}\Big(\widehat P^2_{12}\Big) t_{\rm EW}, \nonumber\\
\Omega&=\frac{1}{2}\sqrt{(\xi_{11}-\xi_{22})^2+4 \xi_{12}^2}.
\end{align}
Using that $\xi_{22}-\xi_{11}$ is proportional to of $y_\tau^2/(16\pi^2)$, with $y_\tau\sim 10^{-2}$ the tau Yukawa coupling in the Standard Model, one can approximate  $|\xi_{12}|\ll |\xi_{11}-\xi_{22}|\ll 1$ and $\Omega\ll 1$. Then:
\begin{align}
m_1(t_{\rm EW})&=e^{(\xi_{11}+\xi_{22})/2}\Big[m_1(0)+ \xi_{12} m_2(0)\Big], \nonumber\\
m_2(t_{\rm EW})&=e^{(\xi_{11}+\xi_{22})/2}\Big[m_2(0)+ \xi_{12} m_1(0)\Big].
\end{align}
The most notable feature in this equation is the dependence of the neutrino masses at the electroweak scale on {\it both} neutrino masses at the cut-off scale. This behavior can be understood from the symmetries in the Lagrangian. The kinetic terms for the left-handed neutrinos possess a $U(2)_L$ global symmetry, which is broken to $U(1)_e\times U(1)_\mu$ by the Yukawa couplings, leading to the conservation of the individual family lepton numbers. As is well known, if at the cut-off scale $m_1(0)=0$ and $m_2(0)=0$, the $U(1)_e\times U(1)_\mu$ symmetry remains unbroken, leading to two massless neutrinos at the electroweak scale. Further, if both $m_1(0)\neq 0$ and $m_2(0)\neq 0$, then the $U(1)_e\times U(1)_\mu$ symmetry is broken, leading to no residual lepton family symmetry or total lepton symmetry at low energies, and to two non-vanishing eigenvalues at all scales. More interesting is the case where $m_1(0)=0$, but $m_2(0)\neq 0$. Then there is a residual global $U(1)$ symmetry if there is no flavor mixing ($\xi_{12}=0$), and no symmetry otherwise. In the former case, one then finds a ``chiral" symmetry for the neutrino mass eigenstate $\nu_1$ that protects the mass against quantum effects, while in the latter case, there is no symmetry preventing $m_1$ from becoming non-zero. This generates a non-vanishing mass for the lightest neutrino, which is proportional to the order parameters of the symmetry breaking $U(1)_e\times U(1)_\mu\rightarrow \emptyset$, namely $m_2(0)$ and $\xi_{12}$. More quantitatively, for the Standard Model values of the Yukawa and gauge couplings, and for the central values of the mixing angles reported in Ref.~\cite{Esteban:2020cvm}, we obtain $(\xi_{11}+\xi_{22})/2\simeq $ $\ln 0.76$ for a cut-off of the theory $\Lambda=10^{14}$ GeV,  of which a factor  $40/6/(16\pi^2) \sim 4\%$ corresponds to the two-loop corrections. Further $\xi_{12}\sim 10^{-11}$. Therefore, the neutrino mass eigenvalues at the electroweak scale are approximately related to their cut-off values by:
\begin{align}
m_1(t_{\rm EW})&\simeq 0.76 \,{\rm max}\big\{m_1(0),  10^{-11} m_2(0)\big\} \;, \nonumber\\
m_2(t_{\rm EW})&\simeq 0.76\, m_2(0) \;,
\end{align}
leading to $m_1(t_{\rm EW}) \sim 10^{-13}$ eV if $m_1(0) = 0$ and $m_2(0) \simeq 0.05$ eV.

In order to understand the implication of the RGEs for the mixing angles and CP violating phases, still in the simplified case of two neutrinos, it is convenient to decompose the PMNS matrix as $U=P^{}_l V P^{}_\nu$ with $P^{}_l = {\rm diag} ( e^{\rm i \phi^{}_e}, e^{\rm i \phi^{}_\mu})$ and $P^{}_\nu = {\rm diag} ( e^{\rm i \rho}, 1 )$ being the diagonal ``unphysical" and ``Majorana" phase matrices respectively. Note that despite the fact that the unphysical phases can be removed at any scale by a proper redefinition of the charged lepton fields, and they do not affect the running of the physical parameters, they are necessary to preserve over the RG evolution the standard parametrization of $V$~\cite{Casas:1999tg}. If all neutrino masses are non-vanishing and much larger than the two-loop corrections, one obtains from Eq.~\eqref{eq:diagonal-rge2} that
\begin{eqnarray}\label{eq:phase-case1}
	\frac{{\rm d}}{{\rm d}t} \left( \phi^{}_e + \phi^{}_\mu + \rho \right) \simeq 0 \;.
\end{eqnarray}
Therefore, in this parametrization, the sum of the unphysical and Majorana phases will be roughly invariant against the RG running. Further, if the two neutrino masses are highly degenerate (namely, when their mass difference at the cut-off is smaller than the splitting induced by the RG running), it follows from Eqs.~\eqref{eq:nondiagonal-rge1} and \eqref{eq:nondiagonal-rge2} that the flavor mixing matrix would be driven  by the RG running to a quasi-fixed point in the infrared, which can be derived from Eq.~(\ref{eq:nondiagonal-rge1}), setting $m_1=m_2=m$,
\begin{align}
	 {\rm Re} \left( \widehat{Q}^{}_{12} +  \widehat{P}^{}_{11} \widehat{P}^{}_{21}+ \widehat{P}^{}_{12} \widehat{P}^{}_{22}\right)=0 \;.
\end{align}

\subsection{Three neutrino case}

Most of the conclusions from the two neutrino case also apply to the three neutrino case, with straightforward modifications. Therefore, here we will focus only on the increase of the rank of the mass matrix and its possible consequences. Using Eqs.~\eqref{eq:diagonal-rge1} and \eqref{eq:diagonal-rge2} one obtains the value of the lightest neutrino mass, as well as a correlation among the CP violating phases, which reads
\begin{eqnarray}\label{eq:mass-NO-case2}
	m^{}_1 \left( \Lambda^{}_{\rm EW} \right) &\sim& \frac{2 y^4_\tau }{\left( 16\pi^2 \right)^2} \ln\left( \frac{\Lambda}{\Lambda^{}_{\rm EW}} \right) \cdot \sum_{i=2,3} m^{}_i \; {\rm Re} \left( U^{\ast}_{\tau 1} U^{}_{\tau i} \right)^2 \;,
	\nonumber \\\label{eq:phase-NO-case2}
	0 &=& m^{}_2 {\rm Im} \left( U^{\ast}_{\tau 1} U^{}_{\tau 2} \right)^2 + m^{}_3 {\rm Im} \left( U^{\ast}_{\tau 1} U^{}_{\tau 3} \right)^2 \;,
\end{eqnarray}
for $m^{}_1 (\Lambda) = 0$ in the normal neutrino mass ordering (NMO) and
\begin{eqnarray}\label{eq:mass-IO-case2}
	m^{}_3 \left( \Lambda^{}_{\rm EW} \right) &\sim& \frac{2 y^4_\tau }{\left( 16\pi^2 \right)^2} \ln\left( \frac{\Lambda}{\Lambda^{}_{\rm EW}} \right) \cdot \sum_{i=1, 2} m^{}_i \; {\rm Re} \left( U^{\ast}_{\tau 3} U^{}_{\tau i} \right)^2 \;,
	\nonumber \\\label{eq:phase-IO-case2}
	0 &=& m^{}_1 {\rm Im} \left( U^{\ast}_{\tau 3} U^{}_{\tau 1} \right)^2 + m^{}_2 {\rm Im} \left( U^{\ast}_{\tau 3} U^{}_{\tau 2} \right)^2 \;,
\end{eqnarray}
for $m^{}_3 (\Lambda) = 0$ in the inverted neutrino mass ordering (IMO). 

To this end, we first to decompose the PMNS matrix as $U=P^{}_l V P^{}_\nu$ with $P^{}_l = {\rm diag}(  e^{\rm i \phi^{}_e}, e^{\rm i \phi^{}_\mu}, e^{\rm i \phi^{}_\tau} )$ and $P^{}_\nu = {\rm diag}( e^{\rm i \rho}, e^{\rm i \sigma}, 1 )$ being the diagonal ``unphysical" and ``Majorana" phase matrices respectively. The ``Dirac" CP-violating phase is contained in the unitary matrix $V$.  If all neutrino masses are non-vanishing, one obtains from   Eq.~\eqref{eq:diagonal-rge2} that 
\begin{eqnarray}\label{eq:phase-case1}
	\frac{{\rm d}}{{\rm d}t} \left( \phi^{}_e + \phi^{}_\mu + \phi^{}_\tau + \rho + \sigma \right) \simeq \rmi\; {\rm Tr} \left( V^\dagger \frac{{\rm d} V}{{\rm d} t} \right) \;.
\end{eqnarray}
which vanishes with the standard parametrization of the PMNS matrix from Ref.~\cite{ParticleDataGroup:2022pth} (this is not necessarily the case for other parametrization~\cite{Xing:2005fw,Ohlsson:2012pg,Zhang:2020lsd, Fritzsch:1997fw}). Therefore, in this parametrization, the sum of the unphysical and Majorana phases will be nearly invariant against the RG running. 

\begin{figure}[t!]
	\centering
	\subfigure[NMO]{
		\includegraphics[width=0.48\linewidth]{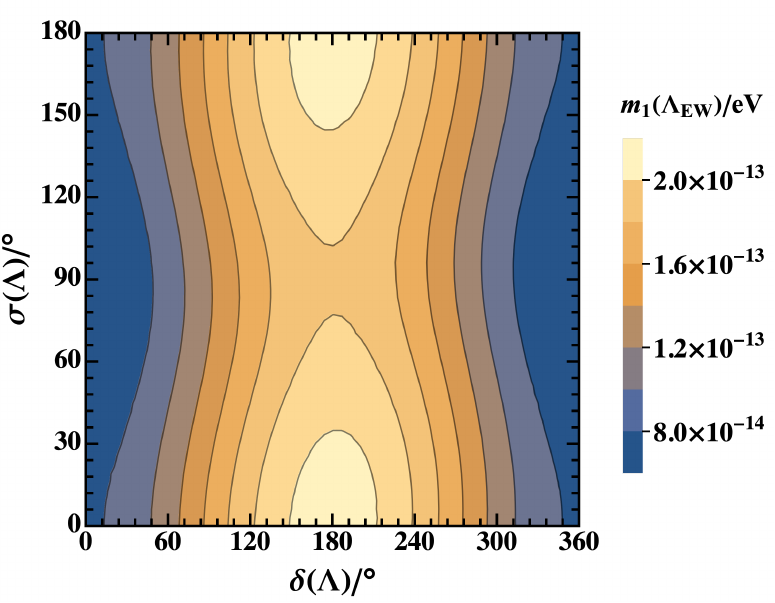}
	}
	\subfigure[IMO]{
		\includegraphics[width=0.48\linewidth]{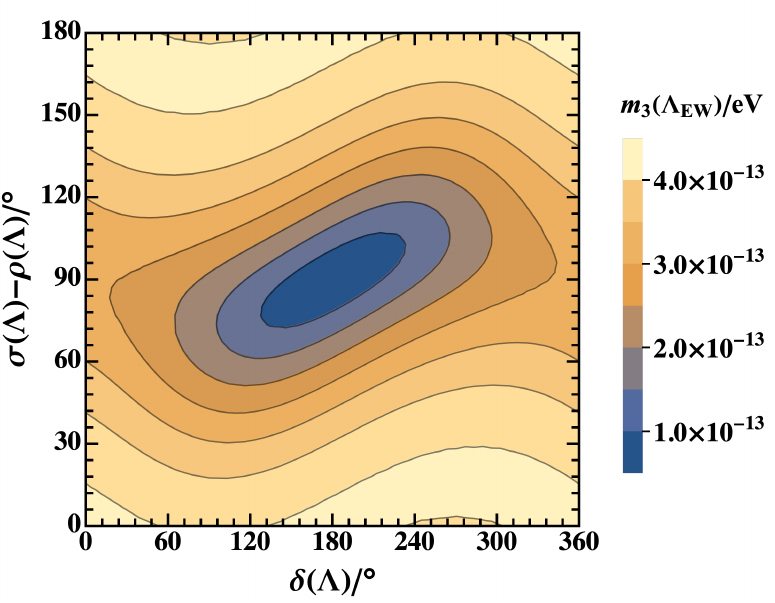}
	}
	\vspace{-0.2cm}
	\caption{Contours of the lightest neutrino mass $m^{}_1 \left( \Lambda^{}_{\rm EW} \right)$ or $m^{}_3 \left( \Lambda^{}_{\rm EW} \right)$ with $\Lambda^{}_{\rm EW} = 200$ GeV  as a function of the two physical phases in the leptonic mixing matrix, the ``Dirac phase" ($\delta$) and the ``Majorana phase" ($\sigma$ for NMO and $\sigma-\rho$ for IMO) at the scale  $\Lambda = 10^{14} $ GeV.}
	\label{fig:lightest-mass}
\end{figure}

We show in Fig. \ref{fig:lightest-mass} contours of the radiatively generated lightest neutrino mass at the electroweak scale $\Lambda^{}_{\rm EW}$ ($m_1$ for NMO and $m_3$ for IMO) as a function of the two physical phases in the leptonic mixing matrix, the ``Dirac phase" ($\delta$) and the ``Majorana phase" ($\sigma$ for NMO and $\sigma-\rho$ for IMO). For the plot we fixed the mass splittings and mixing angles at $\Lambda^{}_{\rm EW} =200$ GeV to the central values reported in NuFIT~5.2~\cite{Esteban:2020cvm}, namely, $\left( \Delta m^2_{21} /{\rm eV}^2, \Delta m^2_{3\ell}/{\rm eV}^2, \theta^{}_{12}/^\circ, \theta^{}_{13}/^\circ, \theta^{}_{23}/^\circ \right)= \left( 7.41\times 10^{-5}, 2.507\times 10^{-3}, 33.41, 8.58, 42.2 \right)$ and $\left( 7.41\times 10^{-5}, -2.486\times 10^{-3}, 33.41, 8.57, 49.0 \right)$ with $\Delta m^2_{3\ell} = \Delta m^2_{31} >0 $ and  $\Delta m^2_{3\ell} =\Delta m^2_{32} < 0$ for the NMO and IMO respectively. One can see from Fig. \ref{fig:lightest-mass} that the generated lightest neutrino mass in the IMO case is slightly larger than that in the NMO case. This is due to the nearly degenerate neutrino masses $m^{}_1 \simeq m^{}_2$ in the former case and the hierarchy among neutrino masses $m^{}_2$ and $m^{}_3$ in the latter case. The lightest neutrino mass has a stronger dependence on the Dirac phase than on the Majorana one in the NMO case, while it is the opposite in the IMO case. This can be traced to the  explicit expressions for $m^{}_1$ or $m^{}_3$, from where it follows the $\sigma$-term is suppressed by $m^{}_2/m^{}_3$ in the NMO case and instead the $\delta$-term is suppressed by $\sin\theta^{}_{13}$ in the IMO case (see also Ref.~\cite{Xing:2020ezi}).

\begin{figure}[t!]
	\centering
        \hspace{-0.6cm}
	\subfigure[NMO]{
		\includegraphics[width=0.45\linewidth]{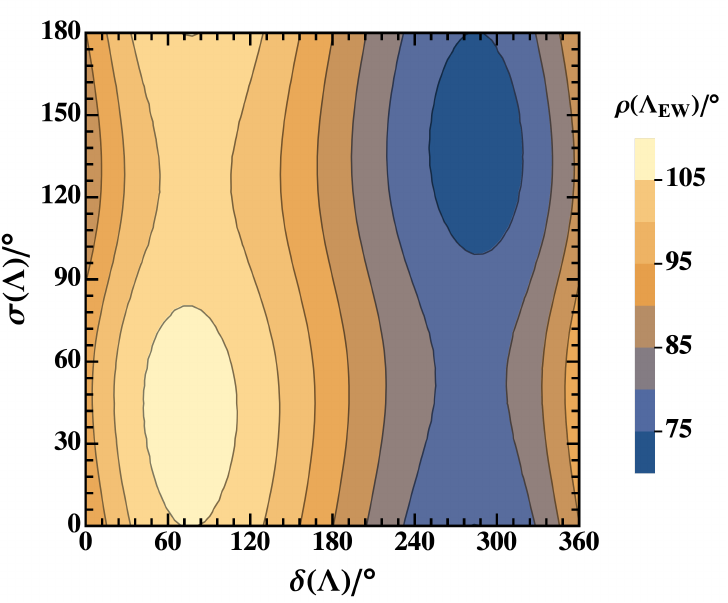}
	}\hspace{0.5cm}
	\subfigure[IMO]{
		\includegraphics[width=0.45\linewidth]{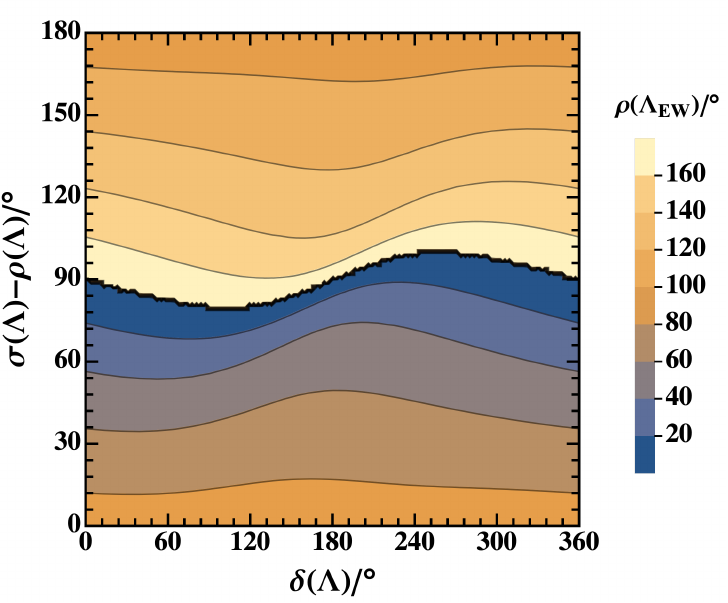}
	}
	\vspace{-0.2cm}
	\caption{Contours of the Majorana phase $\rho\left( \Lambda^{}_{\rm EW} \right)$ associated with the lightest neutrino mass with $\Lambda^{}_{\rm EW} = 200$ GeV in the physical phases $\left[ \delta \left( \Lambda \right), \sigma \left( \Lambda \right) \right]$ or $\left[ \delta \left( \Lambda \right), \sigma \left( \Lambda \right) - \rho \left( \Lambda \right)   \right]$ plane with $\Lambda = 10^{14} $~GeV for the NMO (the left panel) or IMO (the right panel).}
	\label{fig:phase}
\end{figure}

\begin{figure}
	\centering
	\subfigure[NMO]{
		\includegraphics[width=0.48\linewidth]{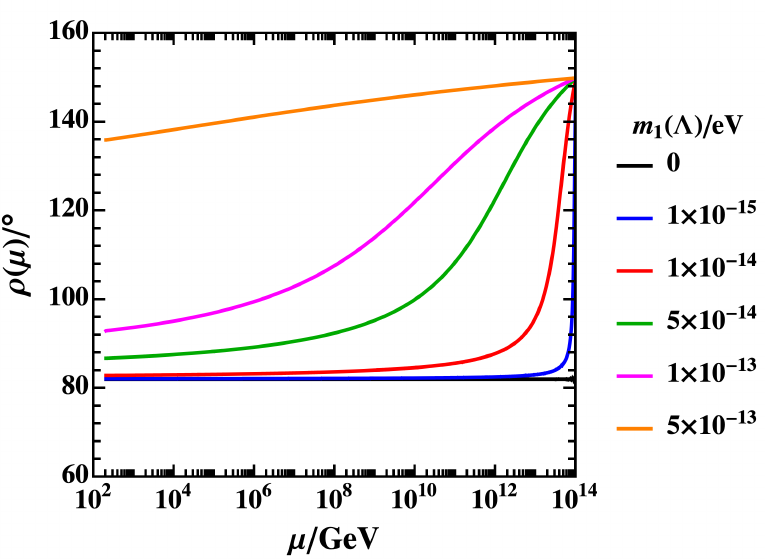}
	}
	\subfigure[IMO]{
		\includegraphics[width=0.48\linewidth]{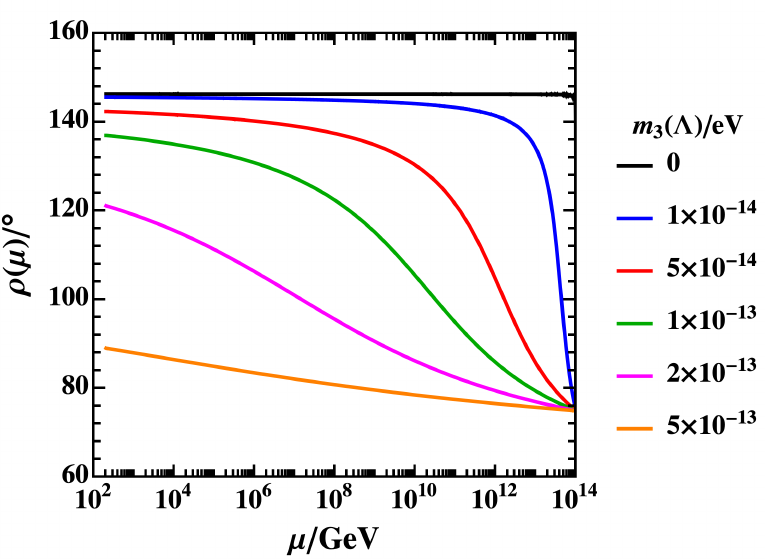}
	}
	\vspace{-0.2cm}
	\caption{Evolution of the Majorana phase $\rho$ from $\Lambda =10^{14}$~GeV to $\Lambda^{}_{\rm EW} = 200$~GeV for different values of the lightest neutrino at the cut-off scale in the NMO (left panel) and IMO (right panel). }
	\label{fig:phase2}
\end{figure}

On the other hand, Eq.~\eqref{eq:phase-NO-case2} results in a correlation among CP-violating phases for the NMO case,
\begin{eqnarray}\label{eq:phase1-NO-case2}
	\tan 2\rho  = - \frac{m^{}_2 \cos2\sigma {\rm Im} \left( V^\ast_{\tau1} V^{}_{\tau2} \right)^2 + m^{}_2 \sin2\sigma {\rm Re} \left( V^\ast_{\tau1} V^{}_{\tau2} \right)^2 + m^{}_3 {\rm Im} \left( V^\ast_{\tau1} V^{}_{\tau3} \right)^2}{m^{}_2 \sin2\sigma {\rm Im} \left( V^\ast_{\tau1} V^{}_{\tau2} \right)^2 - m^{}_2 \cos2\sigma {\rm Re} \left( V^\ast_{\tau1} V^{}_{\tau2} \right)^2 - m^{}_3 {\rm Re} \left( V^\ast_{\tau1} V^{}_{\tau3} \right)^2}
\end{eqnarray}
and  Eq.\eqref{eq:phase-IO-case2} for the IMO case
\begin{eqnarray}\label{eq:phase1-IO-case2}
	\tan 2\rho  = - \frac{ m^{}_1 {\rm Im} \left( V^\ast_{\tau3} V^{}_{\tau1} \right)^2 + m^{}_2 \cos2\left(\sigma - \rho \right) {\rm Im} \left( V^\ast_{\tau3} V^{}_{\tau2} \right)^2 + m^{}_2 \sin2\left( \sigma - \rho \right) {\rm Re} \left( V^\ast_{\tau3} V^{}_{\tau2} \right)^2 }{m^{}_1 {\rm Re} \left( V^\ast_{\tau3} V^{}_{\tau1} \right)^2 - m^{}_2 \sin2\left(\sigma - \rho \right) {\rm Im} \left( V^\ast_{\tau3} V^{}_{\tau2} \right)^2 + m^{}_2 \cos2\left( \sigma - \rho \right) {\rm Re} \left( V^\ast_{\tau3} V^{}_{\tau2} \right)^2}.
\end{eqnarray}
When the lightest neutrino is massless, this phase can be absorbed in the definition of the mass eigenvector. However, this phase becomes physical when the RGEs generate radiatively the lightest neutrino mass, and reaches this quasi-fixed in the infrared, regardless of its value at the cut-off scale. We show in Fig.~\ref{fig:phase} contours of the generated Majorana phase at the electroweak scale $\Lambda^{}_{\rm EW}$. The dependence of $\rho \left( \Lambda^{}_{\rm EW} \right)$ on the physical phases at $\Lambda = 10^{14}$ GeV is very similar to that of the generated lightest neutrino mass and can be well understood from  Eqs.~\eqref{eq:phase1-NO-case2} and \eqref{eq:phase1-IO-case2} for the NMO and IMO respectively. For sizable $m_{1,3}(\Lambda)$, the running only modifies mildly the value of the phase at a low energy scale. However, as $m_{1,3}(\Lambda)\rightarrow 0$ the effect of the RGEs becomes much more significant and drives the value to an infrared fixed point that is given by Eqs.~\eqref{eq:phase1-NO-case2} and \eqref{eq:phase1-IO-case2} for the NMO and IMO cases respectively. This behavior is illustrated in Fig.~\ref{fig:phase2},  where we took $\Lambda = 10^{14}$~GeV, $\delta(\Lambda)= 1.29 \pi$ and $1.53\pi$, and  $\left( \rho, \sigma\right) = \left( 5\pi/6, \pi/6 \right)$ and $\left( 5\pi/12, \pi/12 \right) $  for the NMO and IMO cases, respectively.

Though the generated Majorana phase could be sizable, its effects are necessarily proportional to the lightest neutrino mass. Concretely, from Eqs.~\eqref{eq:mass-NO-case2} and \eqref{eq:phase-IO-case2} one obtains
\begin{eqnarray}
	m^{}_1 e^{2\rmi \rho} \sim \epsilon_\tau \left[ m^{}_2 e^{2\rmi \sigma} \left( V^\ast_{\tau 1} V^{}_{\tau 2} \right)^2 + m^{}_3 \left( V^\ast_{\tau 1} V^{}_{\tau 3} \right)^2  \right]
\end{eqnarray}
for the NMO case and 
\begin{eqnarray}
	m^{}_3 e^{-2\rmi \rho} \sim \epsilon_\tau \left[ m^{}_1 \left( V^\ast_{\tau 3} V^{}_{\tau 1} \right)^2 + m^{}_2 e^{2\rmi \left( \sigma - \rho \right) } \left( V^\ast_{\tau 3} V^{}_{\tau 2} \right)^2  \right]
\end{eqnarray}
for the IMO case. Here, $\epsilon_\tau = 2y^4_\tau \ln\left( \Lambda/\Lambda^{}_{\rm EW} \right) /\left( 16 \pi^2 \right)^2 \sim 10^{-11} $ for $\Lambda = 10^{14}$ GeV. One can then replace this relation into some observable quantity, like the effective neutrino mass for $0\nu\beta\beta$ decay, resulting in: 
\begin{eqnarray}\label{eq:0nbb}
	\left| m^{}_{\beta\beta} \right| = \left\{ \begin{aligned} &\left| m^{}_2 e^{2\rmi \sigma} \left[ V^2_{e2}  + \epsilon_\tau V^2_{e1} \left( V^\ast_{\tau 1} V^{}_{\tau 2} \right)^2 \right] + m^{}_3 \left[ V^2_{e3} + \epsilon_\tau V^2_{e1} \left( V^\ast_{\tau 1} V^{}_{\tau 3} \right)^2  \right] \right| \;, & {\rm NMO} \;, \\[0.3cm] &\left| m^{}_1 \left[ V^2_{e1} + \epsilon_\tau V^2_{e3} \left( V^\ast_{\tau 3} V^{}_{\tau 1} \right)^2 \right] + m^{}_2  e^{2\rmi\left( \sigma - \rho \right)}  \left[ V^2_{e2} + \epsilon_\tau V^2_{e3} \left( V^\ast_{\tau 3} V^{}_{\tau 2} \right)^2  \right]  \right| \;, & {\rm IMO} \;. \end{aligned}\right.
\end{eqnarray}
which does not depend on the phase $\rho$, confirming that despite the phase being sizable, the final effect on observables is very small (proportional to the lightest neutrino mass or to $\epsilon_\tau$).

\section{Conclusions}\label{sec:conclusions}

We have presented the full two-loop RGE of the Weinberg operator with the SM particle content, thus completing the set of two-loop RGEs of the Standard Model Effective Field Theory up to dimension 5. The result was checked by validating the iteration relation between the first and second poles of $\varepsilon$ in the UV divergences. In particular, we have carefully examined the rank-increase contributions to the RGE and found that there is a cancellation among the first poles of $\varepsilon$ induced by the additional diagrams. This confirms the result for the rank-increase terms obtained in previous works. Moreover,  we have discussed the RG running effects at the two-loop level on the lightest neutrino mass and the associated Majorana CP-violating phase, both analytically and numerically. We have calculated a lower limit on the lightest neutrino mass for the normal and the inverted neutrino mass orderings as a function of the Dirac and Majorana phases of the model, which is $\mathcal{O}(10^{-13})$~eV. The Majorana phase, associated with the lightest neutrino mass, reaches a quasi-fixed point in the infrared and can be sizable. However, the smallness of the lightest neutrino mass translates into a very modest impact of this phase in experiments, and concretely in the search for neutrinoless double beta decay.

\begin{acknowledgments} 

NL is supported by the Cluster of Excellence Precision Physics, Fundamental Interactions and Structure of Matter (PRISMA+ EXC 2118/1) funded by the German Research Foundation (DFG) within the German Excellence Strategy (Project No. 390831469). DZ is supported by a Fellowship of the Alexander von Humboldt Foundation. This work was partially supported by the Collaborative Research Center SFB1258 and by the Deutsche Forschungsgemeinschaft (DFG, German Research Foundation) under Germany's Excellence
Strategy - EXC-2094 - 390783311. 
\end{acknowledgments}

\appendix

\section{One-loop Counterterms}\label{app:one-loop}

To determine the RGEs up to two-loops, it is necessary to calculate first the one-loop counterterms, both for determining the one-loop RGEs and for eliminating subdivergences from two-loop diagrams. In particular, working with BFM in Feynman-'t Hooft gauge and the infrared rearrangement approach in Ref.~\cite{Chetyrkin:1997fm}, one has to calculate the one-loop counterterms for gauge-fixing parameters and the spurious masses of the quantum gauge fields and Higgs field from the corresponding self-energy diagrams. These counterterms are not necessary at the one-loop level but are necessary at the two-loop level. The counterterms of the gauge-fixing parameters are necessary for the renormalization of the longitudinal part of the gauge field propagator~\cite{Abbott:1980hw}, while those of spurious masses to cancel the (sub) divergences proportional to $M^2$~\cite{Chetyrkin:1997fm}. We extracted the one-loop UV divergences in terms of the Passarino-Veltman integrals~\cite{tHooft:1978jhc,Passarino:1978jh} (see also Ref.~\cite{Denner:1991kt}), using {\sf FeynCalc}. 

The one-loop counterterms of the different fields and the SM couplings read:
\begin{eqnarray}\label{eq:one-loop-results}
	\delta Z^{(1)}_G &=& \frac{7g^2_3}{16\pi^2 \varepsilon} \;,
	\nonumber
	\\
	\delta Z^{(1)}_W &=& \frac{19g^2_2}{16\pi^2 \varepsilon \cdot 6} \;,
	\nonumber
	\\
	\delta Z^{(1)}_B &=& -\frac{41g^2_1}{16\pi^2 \varepsilon \cdot 6} \;,
	\nonumber
	\\
	\delta Z^{(1)}_{\xi_{\hat{G}}} &=& \frac{g^2_3}{16\pi^2 \varepsilon} \;,
	\nonumber
	\\
	\delta Z^{(1)}_{\xi_{\hat{W}}} &=& -\frac{5g^2_2}{16\pi^2 \varepsilon \cdot 6} \;,
	\nonumber
	\\
	\delta Z^{(1)}_{\xi_{\hat{B}}} &=& - \frac{41g^2_1}{16\pi^2 \varepsilon \cdot 6} \;,
	\nonumber
	\\
	\delta Z^{(1)}_H &=&  \frac{1}{16\pi^2 \varepsilon \cdot 2} \left( g^2_1 + 3g^2_2 - 2T \right) \;,
	\nonumber
	\\
	\delta Z^{(1)}_\lambda &=& \frac{1}{16\pi^2 \varepsilon \cdot 16\lambda} \left[ 32\lambda T + 3 (g1^2+g2^2)^2 + 6g^4_2 - 24\lambda(g^2_1 + 3g^2_2) + 192 \lambda^2 - 16 T^\prime \right] \;,
	\nonumber
	\\
	\delta Z^{(1)}_\ell &=& -\frac{1}{16\pi^2 \varepsilon \cdot 4} \left( g^2_1 + 3g^2_2 + 2 Y^{}_l Y^\dagger_l \right) \;,
	\nonumber
	\\
	\delta Z^{(1)}_e &=& -\frac{1}{16\pi^2 \varepsilon} \left( g^2_1 + Y^\dagger_l Y^{}_l  \right) \;,
	\nonumber
	\\
	\delta Z^{(1)}_q &=& -\frac{1}{16\pi^2 \varepsilon \cdot 36} \left( g^2_1 + 27g^2_2 + 48g^2_3 + 18 Y^{}_{\rm d} Y^\dagger_{\rm d} + 18 Y^{}_{\rm u} Y^\dagger_{\rm u} \right) \;,
	\nonumber
	\\
	\delta Z^{(1)}_u &=& -\frac{1}{16\pi^2 \varepsilon \cdot 9} \left( 4g^2_1 + 12g^2_3 + 9 Y^\dagger_{\rm u} Y^{}_{\rm u} \right) \;,
	\nonumber
	\\
	\delta Z^{(1)}_d &=& -\frac{1}{16\pi^2 \varepsilon \cdot 9} \left( g^2_1 + 12g^2_3 + 9 Y^\dagger_{\rm d} Y^{}_{\rm d} \right) \;,
	\nonumber
	\\
	\delta Z^{(1)}_{Y_l} &=& \frac{1}{16\pi^2 \varepsilon \cdot 8} Y^{-1}_l \left[ Y^{}_l \left( 4T - 15g^2_1 - 9g^2_2 \right)  + 6 Y^{}_l Y^\dagger_l Y^{}_l  \right] \;,
	\nonumber
	\\
	\delta Z^{(1)}_{Y_{\rm u}} &=& \frac{1}{16\pi^2 \varepsilon \cdot 24} Y^{-1}_{\rm u} \left[ Y^{}_{\rm u} \left( 12T - 17g^2_1 - 27g^2_2 - 96 g^2_3 \right)  + 18 Y^{}_{\rm u} Y^\dagger_{\rm u} Y^{}_{\rm u} - 18 Y^{}_{\rm d} Y^\dagger_{\rm d} Y^{}_{\rm u}  \right] \;,
	\nonumber
	\\
	\delta Z^{(1)}_{Y_{\rm d}} &=& \frac{1}{16\pi^2 \varepsilon \cdot 24} Y^{-1}_{\rm d} \left[ Y^{}_{\rm d} \left( 12T - 5g^2_1 - 27g^2_2 - 96 g^2_3 \right)  - 18 Y^{}_{\rm u} Y^\dagger_{\rm u} Y^{}_{\rm d} + 18 Y^{}_{\rm d} Y^\dagger_{\rm d} Y^{}_{\rm d}  \right] \;,
\end{eqnarray}
while those of the spurious gauge field masses read:
\begin{eqnarray}\label{eq:spurious-mass-ct-exp}
        \delta Z^{(1)}_{M_H} &=& \frac{1}{16\pi^2\varepsilon} \left( 6\lambda + T \right) \;,
	\nonumber
	\\
	\delta Z^{(1)}_{M_{\hat{B}}} &=& \frac{1}{16\pi^2\varepsilon} \frac{41}{6} g^2_1 \;,
	\nonumber
	\\
	\delta Z^{(1)}_{M_{\hat{W}}} &=& -\frac{1}{16\pi^2\varepsilon} \frac{7}{6} g^2_2 \;.
\end{eqnarray}
The one-loop counterterm of $C^{}_5$ is calculated in Section \ref{sec:complete} and given in Eq.~\eqref{eq:one-loop-C5-ct}.

\section{Weinberg Operator Vertex Renormalization}\label{app:vertex}

 The Weinberg operator vertex counterterm is directly associated with the four-point loop diagrams and consists of the Weinberg operator's Wilson coefficient counterterm and the wave-function constants of the lepton and Higgs doublets. There are two types of contributions to them. One is from two-loop diagrams, while the other is from one-loop diagrams with an insertion of one-loop counterterms. With the approach described in Section~\ref{sec:complete} the calculation can be reduced to two-loop vacuum diagrams, which were discussed in Refs.~\cite{vanderBij:1983bw,Davydychev:1992mt,Chetyrkin:1997fm}, and the reduction of two-loop tensor integral. For the latter, we used {\sf TARCER}, which is based on the algorithm proposed in Refs.~\cite{Tarasov:1997kx,Tarasov:1996br}. We find only three types of scalar integrals: the vacuum massive one-loop integral and the vacuum two-loop integrals involving one or three massive propagators. Explicitly, they are given by:
\begin{eqnarray}\label{eq:integral}
	A^{(d)}_{\{1,M\}} &=& \frac{1}{\pi^{d/2}} \int \frac{ {\rm d}^d k }{k^2 - M^2} 
	\nonumber
	\\
	&=& -\rmi M^{2-2\varepsilon} \Gamma \left( \varepsilon - 1 \right) \;,
	\nonumber
	\\
	K^{(d)}_{\{1,M\},\{1,0\},\{1,0\}} &=& \frac{1}{\pi^{d}}  \int\int  \frac{ {\rm d}^d k^{}_1 {\rm d}^d k^{}_2 }{\left( k^2_1 -M^2 \right) k^2_2 \left( k^{}_1 - k^{}_2 \right)^2 }
	\nonumber
	\\
	&= & M^{2-4\varepsilon} \frac{1}{1-\varepsilon} \Gamma \left( \varepsilon \right) \Gamma \left( 1- \varepsilon \right) \Gamma \left( 2\varepsilon - 1 \right) \;,
	\nonumber
	\\
	K^{(d)}_{\{1,M\},\{1,M\},\{1,M\}} &=& \frac{1}{\pi^{d}}  \int\int  \frac{ {\rm d}^d k^{}_1 {\rm d}^d k^{}_2 }{\left( k^2_1 -M^2 \right) \left( k^2_2 - M^2 \right) \left[ \left( k^{}_1 - k^{}_2 \right)^2 - M^2 \right] }
	\nonumber
	\\
	&=& -\frac{3 M^2}{2 \varepsilon^2} \left[ 1 + \varepsilon \left( 3 - 2 \gamma^{}_{\rm E} - 2 \ln M^2 \right)   \right] + {\rm ~finite~terms} \;.
\end{eqnarray}
where we have followed the notation of Ref.~\cite{Mertig:1998vk}, and where in the last integral we have shown only the divergent part; the full expression can be found in Refs.~\cite{Davydychev:1992mt,Chetyrkin:1997fm}.
 
Finally, the contribution from the two-loop diagrams to the vertex counterterms is
\begin{eqnarray}
	\delta V^{(2,2)}_5 &=& \frac{1}{\left( 16\pi^2 \right)^2} \frac{1}{\varepsilon^2} \left\{ -\frac{89}{32} g^4_1 C^{}_5 - \frac{85}{32} g^{4}_2 C^{}_5 - \frac{3}{16} g^2_1 g^2_1 C^{}_5 + \lambda  \left( -14 \lambda + g^2_1 + 6g^2_2 \right) C^{}_5 \right.
	\nonumber
	\\
	&& + \left( 2\lambda + \frac{1}{2} T - \frac{13}{8} g^2_1 - \frac{15}{8} g^2_2 \right) \left[ Y^{}_l Y^\dagger_l C^{}_5 + C^{}_5 \left( Y^{}_l Y^\dagger_l \right)^{\rm T} \right]  + \left( T^\prime - 2\lambda T \right) C^{}_5 
	\nonumber
	\\
	&& + \left. \frac{1}{4} \left[ Y^{}_l Y^\dagger_l Y^{}_l Y^\dagger_l C^{}_5 + C^{}_5 \left( Y^{}_l Y^\dagger_l Y^{}_l Y^\dagger_l \right)^{\rm T} \right] - \left( Y^{}_l Y^\dagger_l \right) C^{}_5 \left( Y^{}_l Y^\dagger_l \right)^{\rm T}  \right\}
	\nonumber
	\\
	&& + \frac{1}{\left( 16\pi^2 \right)^2} \frac{1}{\varepsilon} \left\{ \frac{89\Delta - 26}{16} g^4_1 C^{}_5 + \frac{255\Delta - 38}{48} g^{4}_2 C^{}_5 + \frac{3\Delta - 40}{8} g^2_1 g^2_1 C^{}_5 \right.
	\nonumber
	\\
	&& + \frac{1}{4} \lambda  \left[ 8 \left(14\Delta +1 \right) \lambda - \left( 8\Delta + 15 \right) g^2_1 - \left( 48\Delta + 23 \right) g^2_2 \right] C^{}_5 + \left( 4\lambda\Delta -\frac{1}{3} g^2_1 - \frac{5}{6} g^2_2 \right) T C^{}_5 
	\nonumber
	\\
	&& + \left[ - 4 \left( \Delta + 1 \right)\lambda + \frac{1}{2} \left( 1 -2 \Delta \right) T + \frac{26\Delta -33}{8} g^2_1 + \frac{30\Delta - 21}{8} g^2_2 \right] \left[ Y^{}_l Y^\dagger_l C^{}_5 + C^{}_5 \left( Y^{}_l Y^\dagger_l \right)^{\rm T} \right]  
	\nonumber
	\\
	&& + \left( 1 - 2\Delta \right) T^\prime  C^{}_5 +  \frac{1}{4} \left( 7 -2 \Delta \right)  \left[ Y^{}_l Y^\dagger_l Y^{}_l Y^\dagger_l C^{}_5 + C^{}_5 \left( Y^{}_l Y^\dagger_l Y^{}_l Y^\dagger_l \right)^{\rm T} \right] 
	\nonumber
	\\
	&& - \left. \frac{1}{2} \left( 3 - 4\Delta \right) \left( Y^{}_l Y^\dagger_l \right) C^{}_5 \left( Y^{}_l Y^\dagger_l \right)^{\rm T}  \right\}
\end{eqnarray}
while the contributions from the one-loop diagram with one-loop counterterms are 
\begin{eqnarray}
	\delta V^{(2,1.1)}_5 &=& \frac{1}{16\pi^2\varepsilon} \left\{ - \frac{1}{4} \left[ g^2_1 \left( \delta Z^{(1)}_B + 2\delta Z^{(1)}_{\xi_{\hat{B}}} \right) - 3 g^2_2 \left( \delta Z^{(1)}_W -2\delta Z^{(1)}_{\xi_{\hat{W}}} \right) - 8\lambda \delta Z^{(1)}_\lambda \right] C^{}_5 \right.
	\nonumber
	\\
	&& - \frac{1}{2} \left[ \delta Z^{(1)}_\ell Y^{}_l Y^\dagger_l C^{}_5 + C^{}_5 \left( \delta Z^{(1)}_\ell Y^{}_l Y^\dagger_l \right)^{\rm T} - Y^{}_l Y^\dagger_l \delta Z^{(1)}_\ell C^{}_5 - C^{}_5 \left( Y^{}_l Y^\dagger_l \delta Z^{(1)}_\ell \right)^{\rm T} \right.
	\nonumber
	\\
	&& + \left.\left. 2 Y^{}_l \delta Z^{(1)\dagger}_{Y^{}_l} Y^\dagger_l C^{}_5 + 2 C^{}_5 \left( Y^{}_l \delta Z^{(1)\dagger}_{Y^{}_l} Y^\dagger_l \right)^{\rm T} + 2 Y^{}_l \delta Z^{(1)}_{Y^{}_l} Y^\dagger_l C^{}_5 + 2 C^{}_5 \left( Y^{}_l \delta Z^{(1)}_{Y^{}_l} Y^\dagger_l \right)^{\rm T}  \right] \right\}
	\nonumber
	\\
	&& + \frac{1}{16\pi^2} \left\{ \frac{1}{24} g^2_1  \left[ 8\delta Z^{(1)}_{M_H} - 14\delta Z^{(1)}_{M_{\hat{B}}} + \left( 6\Delta - 9 \right) \delta Z^{(1)}_B + \left( 12\Delta + 7 \right)  \delta Z^{(1)}_{\xi_{\hat{B}}} \right] C^{}_5  \right.
	\nonumber
	\\
	&& - \frac{1}{24} g^2_2  \left[ - 20 \delta Z^{(1)}_{M_H} + 2\delta Z^{(1)}_{M_{\hat{W}}} + 9\left( 2\Delta + 1 \right) \delta Z^{(1)}_W -  \left( 36\Delta + 19 \right)  \delta Z^{(1)}_{\xi_{\hat{W}}} \right] C^{}_5 
	\nonumber
	\\
	&& -2 \lambda \left( \delta Z^{(1)}_{M_H}+ \Delta \delta Z^{(1)}_\lambda \right) C^{}_5 + \delta Z^{(1)}_{M_H} \left[  Y^{}_l Y^\dagger_l C^{}_5  + \left( Y^{}_l Y^\dagger_l C^{}_5 \right)^{\rm T} \right] - \frac{1}{2} \left( 1 - \Delta \right)
	\nonumber
	\\
	&& \times \left[ \delta Z^{(1)}_\ell Y^{}_l Y^\dagger_l C^{}_5  + C^{}_5 \left( \delta Z^{(1)}_\ell Y^{}_l Y^\dagger_l \right)^{\rm T}  - Y^{}_l Y^\dagger_l \delta Z^{(1)}_\ell C^{}_5 - C^{}_5 \left( Y^{}_l Y^\dagger_l \delta Z^{(1)}_\ell \right)^{\rm T}  \right.
	\nonumber
	\\
	&& + \left. \left. 2 Y^{}_l \delta Z^{(1)\dagger}_{Y^{}_l} Y^\dagger_l C^{}_5 + 2 C^{}_5 \left( Y^{}_l \delta Z^{(1)\dagger}_{Y^{}_l} Y^\dagger_l \right)^{\rm T} + 2 Y^{}_l \delta Z^{(1)}_{Y^{}_l} Y^\dagger_l C^{}_5 + 2 C^{}_5 \left( Y^{}_l \delta Z^{(1)}_{Y^{}_l} Y^\dagger_l \right)^{\rm T}  \right] \right\} \;\;\;\;
\end{eqnarray}
and
\begin{eqnarray}
	\delta V^{(2,1.2)}_5 &=& \frac{1}{16\pi^2\varepsilon} \left\{ \left( \lambda + \frac{1}{8} g^2_1 - \frac{3}{8} g^2_2 \right) \left( 2 \delta Z^{(1)}_H C^{}_5 + \delta Z^{(1)}_\ell C^{}_5 + C^{}_5 \delta Z^{(1)\rm T}_\ell + 2\delta C^{(1)}_5 \right) \right.
	\nonumber
	\\
	&& - \frac{1}{2} \left[ 2\delta Z^{(1)}_H Y^{}_l Y^\dagger_l C^{}_5 + 2 \delta Z^{(1)}_H C^{}_5 \left( Y^{}_l Y^\dagger_l \right)^{\rm T} + Y^{}_l Y^\dagger_l C^{}_5 \delta Z^{(1)\rm T}_\ell + \delta Z^{(1)}_\ell C^{}_5 \left( Y^{}_l Y^\dagger_l \right)^{\rm T}  \right.
	\nonumber
	\\
	&& + \left.\left. Y^{}_l Y^\dagger_l \delta Z^{(1)}_\ell C^{}_5 + C^{}_5 \left( Y^{}_l Y^\dagger_l\delta Z^{(1)}_\ell \right)^{\rm T} + 2 Y^{}_l Y^\dagger_l \delta C^{(1)}_5 + 2 \delta C^{(1)}_5 \left( Y^{}_l Y^\dagger_l \right)^{\rm T} \right] \right\}
	\nonumber
	\\
	&& + \frac{1}{16\pi^2} \left\{ - \left[ \lambda \Delta + \frac{1}{16} \left( 2\Delta - 3\right) g^2_1 - \frac{3}{16} \left( 2\Delta + 1 \right) g^2_2 \right] \left( 2 \delta Z^{(1)}_H C^{}_5 + \delta Z^{(1)}_\ell C^{}_5 \right.\right.
	\nonumber
	\\
	&& + \left. C^{}_5 \delta Z^{(1)\rm T}_\ell + 2\delta C^{(1)}_5 \right)  - \frac{1}{2} \left( 1 - \Delta \right) \left[ 2\delta Z^{(1)}_H Y^{}_l Y^\dagger_l C^{}_5 + 2 \delta Z^{(1)}_H C^{}_5 \left( Y^{}_l Y^\dagger_l \right)^{\rm T}  \right.
	\nonumber
	\\
	&& + Y^{}_l Y^\dagger_l C^{}_5 \delta Z^{(1)\rm T}_\ell  + \delta Z^{(1)}_\ell C^{}_5 \left( Y^{}_l Y^\dagger_l \right)^{\rm T} + Y^{}_l Y^\dagger_l \delta Z^{(1)}_\ell C^{}_5 + C^{}_5 \left( Y^{}_l Y^\dagger_l\delta Z^{(1)}_\ell \right)^{\rm T}
	\nonumber
	\\
	&& + \left.\left. 2 Y^{}_l Y^\dagger_l \delta C^{(1)}_5 + 2 \delta C^{(1)}_5 \left( Y^{}_l Y^\dagger_l \right)^{\rm T} \right] \right\} \;.
\end{eqnarray}
Here, $\Delta =  \gamma^{}_{\rm E} - \ln (4\pi) + \ln\left( M^2/\mu^2 \right) $.

Finally, substituting the one-loop counterterms with their explicit results given in Eqs.~\eqref{eq:one-loop-C5-ct}, \eqref{eq:one-loop-results} and~\eqref{eq:spurious-mass-ct-exp} and collecting all contributions, one obtains the two-loop Weinberg operator vertex counterterm, which is given by:
\begin{eqnarray}
	\delta V^{(2)}_5 &=& \frac{1}{\left( 16\pi^2 \right)^2 \varepsilon^2} \left\{ \left[ \frac{89}{32} g^4_1 + \frac{3}{16} g^2_1 g^2_2 + \frac{85}{32} g^4_2 - \lambda \left( g^2_1 + 6g^2_2 - 14\lambda - 2T \right) - T^\prime \right] C^{}_5 \right.
	\nonumber
	\\
	&& + \left( \frac{13}{8} g^2_1 + \frac{15}{8} g^2_2 - 2\lambda - \frac{1}{2} T \right) \left[ Y^{}_l Y^\dagger_l C^{}_5 + C^{}_5 \left( Y^{}_l Y^\dagger_l \right)^{\rm T} \right] - \frac{1}{4} Y^{}_l Y^\dagger_l Y^{}_l Y^\dagger_l C^{}_5 
	\nonumber
	\\
	&& - \left. \frac{1}{4} C^{}_5 \left[ Y^{}_l Y^\dagger_l Y^{}_l Y^\dagger_l  \right]^{\rm T} + Y^{}_l Y^\dagger_l C^{}_5 \left( Y^{}_l Y^\dagger_l \right)^{\rm T} \right\}
	\nonumber
	\\
	&& - \frac{1}{\left( 16\pi^2 \right)^2 \varepsilon} \left\{  \left[ \frac{475}{96} g^4_1 + \frac{83}{16} g^2_1 g^2_2 + \frac{271}{96} g^4_2 + \lambda \left( g^2_1 + 10 \lambda + 2T \right)  - T^\prime \right] C^{}_5 \right.
	\nonumber
	\\
	&& + \left( g^2_1 -  \frac{1}{2} T \right) \left[ Y^{}_l Y^\dagger_l C^{}_5 + C^{}_5 \left( Y^{}_l Y^\dagger_l \right)^{\rm T} \right] - \frac{5}{4} Y^{}_l Y^\dagger_l Y^{}_l Y^\dagger_l C^{}_5 - \frac{5}{4} C^{}_5 \left( Y^{}_l Y^\dagger_l Y^{}_l Y^\dagger_l  \right)^{\rm T}
	\nonumber
	\\
	&& - \left. \frac{1}{2} Y^{}_l Y^\dagger_l C^{}_5 \left( Y^{}_l Y^\dagger_l \right)^{\rm T} \right\} \;,
\end{eqnarray}
which is as expected independent of $\Delta$.

\section{Two-loop Wave-function Renormalization Constants of Lepton and Higgs Doublets} \label{app:lepton-Higgs}

 For two-loop wave-function renormalization constants of lepton and Higgs doublets, there are also two types of contributions from two-loop self-energy diagrams and one-loop ones with one-loop counterterms. The sum of these two kinds of contributions is independent of (spurious) mass parameters. As a result, we can ignore all quadratic terms, including the corresponding counterterms, for simplification when calculating those two types of contributions. Moreover, one- and two-loop self-energy diagrams with an external momentum do not contain any infrared divergence. The one-loop self-energy with one-loop counterterms can be simply calculated, and the two-loop ones with massless propagators are more complicated but can still be analytically calculated via the Feynman parametrization and the integration by parts~\cite{Chetyrkin:1981qh} (see also Ref.~\cite{Grozin:2005yg} for calculation details of such two-loop diagrams). Fortunately, {\sf TARCER} can reduce such two-loop integrals and get their analytical results as the $\varepsilon$ expansion series. Thus, we use {\sf TARCER} to reduce such two-loop integrals and extract their UV divergences. 

The contributions to lepton and Higgs doublet wave-functions renormalization constants from two-loop diagrams are:
\begin{eqnarray}\label{eq:lepton1}
	\delta Z^{(2,2)}_\ell &=& \frac{1}{\left( 16\pi^2 \right)^2 \varepsilon^2 \cdot 32} \left[ - g^4_1 -6g^2_1 g^2_2 - 57 g^4_2  - 2 \left( 17 g^2_1 + 15g^2_2 \right) Y^{}_l Y^\dagger_l + 8 T Y^{}_l Y^\dagger_l + 8 Y^{}_l Y^\dagger_l Y^{}_l Y^\dagger_l \right]
	\nonumber
	\\
	&& +  \frac{1}{\left( 16\pi^2 \right)^2  \varepsilon \cdot 32} \left\{ \frac{1}{2} \left[ \left( 4\Delta^\prime + 81 \right) g^4_1 + 6 \left( 4\Delta^\prime  - 1 \right) g^2_1 g^2_2 + 3\left( 76\Delta^\prime  - 153 \right) g^4_2 \right]  \right.
	\nonumber
	\\
	&& + \left[ \left( 68\Delta^\prime  - 139 \right) g^2_1 + 3 \left( 20\Delta^\prime  - 47 \right)g^2_2 \right] Y^{}_l Y^\dagger_l -4 \left( 4\Delta^\prime  - 11 \right) T Y^{}_l Y^\dagger_l 
	\nonumber
	\\
	&& + \left. 4 \left( 9 -4\Delta^\prime  \right) Y^{}_l Y^\dagger_l Y^{}_l Y^\dagger_l \right\} ,\nonumber \\
	\delta Z^{(2,2)}_{H} &=&  \frac{1}{\left( 16\pi^2 \right)^2  \varepsilon^2} \left\{- \frac{43}{16}g^4_1 - \frac{3}{4} g^2_1 g^2_2 + \frac{15}{16} g^4_2 + \frac{1}{8} g^2_1 {\rm Tr} \left( 7Y^{}_{\rm d} Y^\dagger_{\rm d} - 5Y^{}_{\rm u} Y^\dagger_{\rm u} - 11Y^{}_l Y^\dagger_l \right) + \frac{3}{8} g^2_2 T \right.
	\nonumber
	\\
	&& - \left. 12g^2_3 {\rm Tr} \left( Y^{}_{\rm d} Y^\dagger_{\rm d} + Y^{}_{\rm u} Y^\dagger_{\rm u}  \right) + \frac{3}{4} {\rm Tr} \left( Y^{}_l Y^\dagger_l Y^{}_l Y^\dagger_l + 3 Y^{}_{\rm u} Y^\dagger_{\rm u} Y^{}_{\rm u} Y^\dagger_{\rm u} + 3 Y^{}_{\rm d} Y^\dagger_{\rm d} Y^{}_{\rm d} Y^\dagger_{\rm d} -  6 Y^{}_{\rm u} Y^\dagger_{\rm u} Y^{}_{\rm d} Y^\dagger_{\rm d} \right)  \right\} 
	\nonumber
	\\
	&& +  \frac{1}{\left( 16\pi^2 \right)^2 \varepsilon} \left\{ \left( \frac{43}{8}\Delta^\prime - \frac{613}{64} \right) g^4_1 + \left( \frac{3}{2} \Delta^\prime - \frac{105}{32} \right) g^2_1 g^2_2 - \left(  \frac{15}{8}\Delta^\prime - \frac{483}{64} \right) g^4_2 - 3\lambda^2 \right.
	\nonumber
	\\
	&& + \frac{1}{48} g^2_1 {\rm Tr} \left[ \left( 143 - 84\Delta^\prime \right)Y^{}_{\rm d} Y^\dagger_{\rm d} + 5\left( 12\Delta^\prime - 41 \right) Y^{}_{\rm u} Y^\dagger_{\rm u} + 3\left( 44\Delta^\prime - 113 \right) Y^{}_l Y^\dagger_l \right] 
	\nonumber
	\\
	&&  - \frac{3}{16} \left( 4\Delta^\prime - 3 \right) g^2_2 T + 2 \left( 12\Delta^\prime - 29 \right) g^2_3 {\rm Tr} \left( Y^{}_{\rm d} Y^\dagger_{\rm d} + Y^{}_{\rm u} Y^\dagger_{\rm u}  \right) +   \frac{3}{8} {\rm Tr} \left[ \left( 11 - 4\Delta^\prime \right) \right.
	\nonumber
	\\
	&& \times \left.\left. \left( Y^{}_l Y^\dagger_l Y^{}_l Y^\dagger_l + 3 Y^{}_{\rm u} Y^\dagger_{\rm u} Y^{}_{\rm u} Y^\dagger_{\rm u} + 3 Y^{}_{\rm d} Y^\dagger_{\rm d} Y^{}_{\rm d} Y^\dagger_{\rm d} \right) -  2\left( 25 - 12\Delta^\prime \right) Y^{}_{\rm u} Y^\dagger_{\rm u} Y^{}_{\rm d} Y^\dagger_{\rm d} \right]  \right\} \;,
\end{eqnarray}
and those from one-loop diagrams with one-loop counterterms are
\begin{eqnarray}\label{eq:lepton2}
	\delta Z^{(2,1)}_\ell &=&  \frac{1}{16\pi^2 \varepsilon \cdot 4} \left\{ g^2_1 \left( \delta Z^{(1)}_B - \delta Z^{(1)}_{\xi_{\hat{B}}} \right) + 3g^2_2 \left( \delta Z^{(1)}_W - \delta Z^{(1)}_{\xi_{\hat{W}}} \right) - \left( g^2_1 + 3g^2_2 \right) \delta Z^{(1)}_\ell \right.
	\nonumber
	\\
	&& - \left.  \left( \delta Z^{(1)}_\ell Y^{}_l Y^\dagger_l + Y^{}_l Y^\dagger_l \delta Z^{(1)}_\ell + 2 Y^{}_l \delta Z^{(1)}_{Y^{}_l} Y^\dagger_l + 2 Y^{}_l \delta Z^{(1)\dagger}_{Y^{}_l} Y^\dagger_l \right)  \right\}
	\nonumber
	\\
	&& + \frac{1}{16\pi^2  \cdot 4} \left\{ \left( 1 - \Delta^\prime  \right) \left[ g^2_1 \left( \delta Z^{(1)}_B - \delta Z^{(1)}_{\xi_{\hat{B}}} \right) + 3g^2_2 \left( \delta Z^{(1)}_W - \delta Z^{(1)}_{\xi_{\hat{W}}} \right) - \left( g^2_1 + 3g^2_2 \right) \delta Z^{(1)}_\ell \right] \right.
	\nonumber
	\\
	&& - \left.  \left( 2 - \Delta^\prime  \right) \left( \delta Z^{(1)}_\ell Y^{}_l Y^\dagger_l + Y^{}_l Y^\dagger_l \delta Z^{(1)}_\ell + 2 Y^{}_l \delta Z^{(1)}_{Y^{}_l} Y^\dagger_l + 2 Y^{}_l \delta Z^{(1)\dagger}_{Y^{}_l} Y^\dagger_l \right)  \right\} \;
 \nonumber
 \\
 \delta Z^{(2,1)}_{H} &=& \frac{1}{16\pi^2 \varepsilon} \left\{ -\frac{1}{4} g^2_1 \left( 2\delta Z^{(1)}_B + \delta Z^{(1)}_{\xi_{\hat{B}}} - 2\delta Z^{(1)}_H \right) - \frac{3}{4} g^2_2 \left( 2\delta Z^{(1)}_W + \delta Z^{(1)}_{\xi_{\hat{W}}} - 2\delta Z^{(1)}_H \right) - T \delta Z^{(1)}_H \right.
	\nonumber
	\\
	&& - \left. {\rm Tr} \left[ Y^{}_l \left( \delta Z^{(1)}_{Y^{}_l} + \delta Z^{(1)\dagger}_{Y^{}_l} \right) Y^\dagger_l  + 3Y^{}_{\rm u} \left( \delta Z^{(1)}_{Y^{}_{\rm u}} + \delta Z^{(1)\dagger}_{Y^{}_{\rm u}} \right) Y^\dagger_{\rm u}  + 3Y^{}_{\rm d} \left( \delta Z^{(1)}_{Y^{}_{\rm d}} + \delta Z^{(1)\dagger}_{Y^{}_{\rm d}} \right) Y^\dagger_{\rm d}  \right] \right\}
	\nonumber
	\\
	&& +  \frac{1}{16\pi^2 } \left\{ \frac{1}{4} g^2_1 \left[ 2\left( \Delta^\prime -2 \right) \delta Z^{(1)}_B + \Delta^\prime \delta Z^{(1)}_{\xi_{\hat{B}}} - 2\left( \Delta^\prime - 2\right) \delta Z^{(1)}_H \right] + \frac{3}{4} g^2_2 \left[ 2\left( \Delta^\prime -2 \right) \delta Z^{(1)}_W  \right.\right.
	\nonumber
	\\
	&& + \left. \Delta^\prime \delta Z^{(1)}_{\xi_{\hat{W}}} - 2\left( \Delta^\prime - 2 \right) \delta Z^{(1)}_H \right] + \left( \Delta^\prime - 2 \right) T \delta Z^{(1)}_H +  \left( \Delta^\prime - 2 \right) {\rm Tr} \left[ Y^{}_l \left( \delta Z^{(1)}_{Y^{}_l} + \delta Z^{(1)\dagger}_{Y^{}_l} \right) Y^\dagger_l  \right.
	\nonumber
	\\
	&& + \left. \left. 3Y^{}_{\rm u} \left( \delta Z^{(1)}_{Y^{}_{\rm u}} + \delta Z^{(1)\dagger}_{Y^{}_{\rm u}} \right) Y^\dagger_{\rm u}  + 3Y^{}_{\rm d} \left( \delta Z^{(1)}_{Y^{}_{\rm d}} + \delta Z^{(1)\dagger}_{Y^{}_{\rm d}} \right) Y^\dagger_{\rm d}  \right] \right\} \;.
\end{eqnarray}
where $\Delta^\prime = \gamma^{}_{\rm E} - \ln \left( 4\pi \right) + \ln \left( p^2/\mu^2 \right)$ and $p$ is the external momentum. Combining Eqs.~\eqref{eq:lepton1}, \eqref{eq:lepton2} and the one-loop counterterms in Eq.~\eqref{eq:one-loop-results}, one finally finds the two-loop wave-function renormalization constants:
\begin{eqnarray}
	\delta Z^{(2)}_\ell &=& \frac{1}{\left( 16\pi^2 \right)^2 \varepsilon^2 \cdot 32} \left[  g^4_1 + 6g^2_1 g^2_2 + 57 g^4_2  + 2 \left( 17 g^2_1 + 15g^2_2 \right) Y^{}_l Y^\dagger_l - 8 T Y^{}_l Y^\dagger_l - 8 Y^{}_l Y^\dagger_l Y^{}_l Y^\dagger_l \right]
	\nonumber
	\\
	&& +  \frac{1}{\left( 16\pi^2 \right)^2 \varepsilon \cdot 64} \left[  85g^4_1 + 18 g^2_1 g^2_2 - 231 g^4_2  - 2 \left( 7 g^2_1 + 33 g^2_2 \right) Y^{}_l Y^\dagger_l + 24 T Y^{}_l Y^\dagger_l \right.
	\nonumber
	\\
	&& + \left. 8 Y^{}_l Y^\dagger_l Y^{}_l Y^\dagger_l \right] \;,\\
	\delta Z^{(2)}_H &=&  \frac{1}{\left( 16\pi^2 \right)^2  \varepsilon^2} \left\{ \frac{43}{16}g^4_1 + \frac{3}{4} g^2_1 g^2_2 - \frac{15}{16} g^4_2 + \frac{1}{8} g^2_1 {\rm Tr} \left( 11Y^{}_l Y^\dagger_l + 5Y^{}_{\rm u} Y^\dagger_{\rm u} - 7Y^{}_{\rm d} Y^\dagger_{\rm d}  \right) - \frac{3}{8} g^2_2 T \right.
	\nonumber
	\\
	&& + \left. 12g^2_3 {\rm Tr} \left( Y^{}_{\rm d} Y^\dagger_{\rm d} + Y^{}_{\rm u} Y^\dagger_{\rm u}  \right) - \frac{3}{4} {\rm Tr} \left( Y^{}_l Y^\dagger_l Y^{}_l Y^\dagger_l + 3 Y^{}_{\rm u} Y^\dagger_{\rm u} Y^{}_{\rm u} Y^\dagger_{\rm u} + 3 Y^{}_{\rm d} Y^\dagger_{\rm d} Y^{}_{\rm d} Y^\dagger_{\rm d} -  6 Y^{}_{\rm u} Y^\dagger_{\rm u} Y^{}_{\rm d} Y^\dagger_{\rm d} \right)  \right\} 
	\nonumber
	\\
	&& +  \frac{1}{\left( 16\pi^2 \right)^2 \varepsilon} \left\{ -\frac{431}{192}g^4_1 - \frac{9}{32} g^2_1 g^2_2 + \frac{163}{64} g^4_2 - 3\lambda^2 - \frac{5}{48} g^2_1 {\rm Tr} \left( 15Y^{}_l Y^\dagger_l + 17Y^{}_{\rm u} Y^\dagger_{\rm u} + 5Y^{}_{\rm d} Y^\dagger_{\rm d}  \right)  \right.
	\nonumber
	\\
	&& -  \frac{15}{16} g^2_2 T + \frac{3}{8} {\rm Tr} \left( 3Y^{}_l Y^\dagger_l Y^{}_l Y^\dagger_l + 9 Y^{}_{\rm u} Y^\dagger_{\rm u} Y^{}_{\rm u} Y^\dagger_{\rm u} + 9 Y^{}_{\rm d} Y^\dagger_{\rm d} Y^{}_{\rm d} Y^\dagger_{\rm d} -  2 Y^{}_{\rm u} Y^\dagger_{\rm u} Y^{}_{\rm d} Y^\dagger_{\rm d} \right) 
	\nonumber
	\\
	&& - \left.10 g^2_3 {\rm Tr} \left( Y^{}_{\rm d} Y^\dagger_{\rm d} + Y^{}_{\rm u} Y^\dagger_{\rm u}  \right)   \right\}  \;,
\end{eqnarray}
which are as expected independent of $\Delta^\prime$.

\section{The SM Two-loop RGEs}\label{app:sm}

\begin{eqnarray}
	\mu \frac{{\rm d} g^{}_1}{{\rm d} \mu} &=& \frac{1}{16\pi^2} g^3_1  \left\{ \frac{41}{6}  + \frac{1}{16\pi^2 }  \left[ \frac{199}{18} g^2_1 + \frac{9}{2} g^2_2 + \frac{44}{3} g^2_3 - \frac{1}{6} {\rm Tr} \left( 15 Y^{}_l Y^\dagger_l + 17 Y^{}_{\rm u} Y^\dagger_{\rm u} + 5 Y^{}_{\rm d} Y^\dagger_{\rm d} \right) \right] \right\}\;,
	\nonumber
	\\
	\mu \frac{{\rm d} g^{}_2}{{\rm d} \mu} &=& - \frac{1}{16\pi^2} g^3_2 \left[ \frac{19}{6} - \frac{1}{16\pi^2 } \left( \frac{3}{2} g^2_1 + \frac{35}{6} g^2_2 + 12 g^2_3 - \frac{1}{2}  T \right) \right] \;,
	\nonumber
	\\
	\mu \frac{{\rm d} g^{}_3}{{\rm d} \mu} &=& - \frac{1}{16\pi^2} g^3_3 \left\{ 7 -  \frac{1}{16\pi^2 }  \left[ \frac{11}{6} g^2_1 + \frac{9}{2} g^2_2 - 26 g^2_3 - 2 {\rm Tr} \left(  Y^{}_{\rm u} Y^\dagger_{\rm u} + Y^{}_{\rm d} Y^\dagger_{\rm d} \right) \right]  \right\} \;,
	\nonumber
	\\
	\mu \frac{{\rm d} M^2}{{\rm d} \mu} &=&  \frac{1}{16\pi^2} M^2 \left\{ -\frac{3}{2} g^2_1 - \frac{9}{2} g^2_2 + 12 \lambda + 2 T +  \frac{1}{16\pi^2} \left[ \frac{557}{48} g^4_1 - \frac{145}{16} g^4_2 + \frac{15}{8} g^2_1 g^2_2 \right.\right.
	\nonumber
	\\
	&&
	+ \frac{5}{12} g^2_1 {\rm Tr} \left( 15 Y^{}_l Y^\dagger_l + 17 Y^{}_{\rm u} Y^\dagger_{\rm u} + 5 Y^{}_{\rm d} Y^\dagger_{\rm d} \right) + \frac{15}{4} g^2_2 T + 40 g^2_3 {\rm Tr} \left(  Y^{}_{\rm u} Y^\dagger_{\rm u} + Y^{}_{\rm d} Y^\dagger_{\rm d} \right) 
	\nonumber
	\\
	&& - \left. \left.  \frac{9}{2} T^\prime - 21 {\rm Tr} \left( Y^{}_{\rm u} Y^\dagger_{\rm u} Y^{}_{\rm d} Y^\dagger_{\rm d} \right) + 2\lambda \left( 12 g^2_1 + 36 g^2_2 - 30\lambda - 12 T \right)  \right]  \right\} \;,
	\nonumber
	\\
	\mu \frac{{\rm d} \lambda}{{\rm d} \mu} &=& \frac{1}{16\pi^2} \left[ \frac{3}{8} g^4_1 + \frac{9}{8} g^4_2 + \frac{3}{4} g^2_1 g^2_2 - 2 T^\prime + \left( -3g^2_1 - 9g^2_2 + 24\lambda + 4T \right) \lambda \right] + \frac{1}{\left( 16 \pi^2 \right)^2} \left\{ - \frac{379}{48} g^6_1 \right.
	\nonumber
	\\
	&&  + \frac{305}{16} g^6_2 - \frac{289}{48} g^2_1 g^4_2 - \frac{559}{48} g^4_1 g^2_2 - \frac{1}{4} g^4_1 {\rm Tr} \left( 25 Y^{}_l Y^\dagger_l + 19 Y^{}_{\rm u} Y^\dagger_{\rm u} - 5 Y^{}_{\rm d} Y^\dagger_{\rm d} \right) - \frac{3}{4} g^4_2 T
	\nonumber
	\\
	&&  + \frac{1}{2} g^2_1 g^2_2 {\rm Tr} \left( 11Y^{}_l Y^\dagger_l + 21 Y^{}_{\rm u} Y^\dagger_{\rm u} + 9 Y^{}_{\rm d} Y^\dagger_{\rm d} \right) - \frac{4}{3} g^2_1 {\rm Tr} \left( 3Y^{}_l Y^\dagger_l Y^{}_l Y^\dagger_l + 2 Y^{}_{\rm u} Y^\dagger_{\rm u} Y^{}_{\rm u} Y^\dagger_{\rm u} \right.
	\nonumber
	\\
	&& - \left.  Y^{}_{\rm d} Y^\dagger_{\rm d} Y^{}_{\rm d} Y^\dagger_{\rm d}  \right) - 32 g^2_3 {\rm Tr} \left( Y^{}_{\rm u} Y^\dagger_{\rm u} Y^{}_{\rm u} Y^\dagger_{\rm u} + Y^{}_{\rm d} Y^\dagger_{\rm d} Y^{}_{\rm d} Y^\dagger_{\rm d}  \right) + \lambda \left[ \frac{629}{24} g^4_1 - \frac{73}{8} g^4_2 + \frac{39}{4} g^2_1 g^2_2 \right.
	\nonumber
	\\
	&& +  \frac{5}{6} g^2_1 {\rm Tr} \left( 15 Y^{}_l Y^\dagger_l + 17 Y^{}_{\rm u} Y^\dagger_{\rm u} + 5 Y^{}_{\rm d} Y^\dagger_{\rm d} \right) + \frac{15}{2} g^2_2 T + 80g^2_3 {\rm Tr} \left( Y^{}_{\rm u} Y^\dagger_{\rm u} + Y^{}_{\rm d} Y^\dagger_{\rm d} \right) - T^\prime 
	\nonumber
	\\
	&& - \left. 42 {\rm Tr} \left( Y^{}_{\rm u} Y^\dagger_{\rm u} Y^{}_{\rm d} Y^\dagger_{\rm d} \right)  \right]  + 2 \lambda^2 \left( 18 g^2_1 + 54 g^2_2 - 156\lambda - 24 T \right) + 2 {\rm Tr} \left( 5 Y^{}_l Y^\dagger_l Y^{}_l Y^\dagger_l Y^{}_l Y^\dagger_l  \right.
	\nonumber
	\\
	&& + \left.\left. 15 Y^{}_{\rm u} Y^\dagger_{\rm u} Y^{}_{\rm u} Y^\dagger_{\rm u} Y^{}_{\rm u } Y^\dagger_{\rm u} + 15 Y^{}_{\rm d} Y^\dagger_{\rm d} Y^{}_{\rm d} Y^\dagger_{\rm d} Y^{}_{\rm d } Y^\dagger_{\rm d} - 3  Y^{}_{\rm d} Y^\dagger_{\rm d} Y^{}_{\rm d} Y^\dagger_{\rm d} Y^{}_{\rm u } Y^\dagger_{\rm u} - 3 Y^{}_{\rm d} Y^\dagger_{\rm d} Y^{}_{\rm u} Y^\dagger_{\rm u} Y^{}_{\rm u } Y^\dagger_{\rm u} \right)  \right\} \;,
	\nonumber
	\\
	\mu \frac{{\rm d} Y^{}_l}{{\rm d} \mu} &=& \frac{1}{16\pi^2} \left\{ - \frac{15}{4} g^2_1 - \frac{9}{4} g^2_2 + T + \frac{3}{2} Y^{}_l Y^\dagger_l + \frac{1}{16\pi^2} \left[ \frac{457}{24} g^4_1 + \frac{9}{4} g^2_1 g^2_2 - \frac{23}{4} g^4_2  + 6 \lambda^2 \right.\right.
	\nonumber
	\\
	&& + \frac{5}{24} g^2_1 {\rm Tr} \left( 15 Y^{}_l Y^\dagger_l + 17 Y^{}_{\rm u} Y^\dagger_{\rm u} + 5 Y^{}_{\rm d} Y^\dagger_{\rm d}  \right)  + \frac{15}{8} g^2_2 T + 20 g^2_3 {\rm Tr} \left( Y^{}_{\rm u} Y^\dagger_{\rm u} + Y^{}_{\rm d} Y^\dagger_{\rm d} \right)- \frac{9}{4} T^\prime 
	\nonumber
	\\
	&& +  \left. \left.  \frac{3}{2} {\rm Tr} \left( Y^{}_{\rm u} Y^\dagger_{\rm u} Y^{}_{\rm d} Y^\dagger_{\rm d} \right) + \left( \frac{129}{16} g^2_1 + \frac{135}{16} g^2_2 - 12\lambda - \frac{9}{4} T \right) Y^{}_l Y^\dagger_l + \frac{3}{2} Y^{}_l Y^\dagger_l Y^{}_l Y^\dagger_l \right] \right\} Y^{}_l \;,
	\nonumber
	\\
	\mu \frac{{\rm d} Y^{}_{\rm u}}{{\rm d} \mu} &=& \frac{1}{16\pi^2} \left\{ - \frac{17}{12} g^2_1 - \frac{9}{4}g^2_2 - 8g^2_3 + T + \frac{3}{2} \left( Y^{}_{\rm u} Y^\dagger_{\rm u} - Y^{}_{\rm d} Y^\dagger_{\rm d} \right)  + \frac{1}{16\pi^2} \left[ \frac{1187}{216} g^4_1 - \frac{23}{4} g^4_2 - 108 g^4_3  \right. \right.
	\nonumber
	\\
	&& - \frac{3}{4} g^2_1 g^2_2 + \frac{19}{9} g^2_1 g^2_3 + 9 g^2_2 g^2_3  + \frac{5}{24} g^2_1 {\rm Tr} \left( 15 Y^{}_l Y^\dagger_l + 17 Y^{}_{\rm u} Y^\dagger_{\rm u} + 5 Y^{}_{\rm d} Y^\dagger_{\rm d}  \right)  + \frac{15}{8} g^2_2 T  
	\nonumber
	\\
	&& + 20 g^2_3 {\rm Tr} \left( Y^{}_{\rm u} Y^\dagger_{\rm u}  + Y^{}_{\rm d} Y^\dagger_{\rm d} \right) + 6 \lambda^2 - \frac{9}{4} T^\prime +  \frac{3}{2} {\rm Tr} \left( Y^{}_{\rm u} Y^\dagger_{\rm u} Y^{}_{\rm d} Y^\dagger_{\rm d} \right) + \left( \frac{223}{48} g^2_1 + \frac{135}{16} g^2_2 + 16 g^2_3 \right.
	\nonumber
	\\
	&& - \left. 12\lambda - \frac{9}{4} T \right) Y^{}_{\rm u} Y^\dagger_{\rm u} + \left( -\frac{43}{48} g^2_1 + \frac{9}{16} g^2_2 - 16 g^2_3 + \frac{5}{4} T \right) Y^{}_{\rm d} Y^\dagger_{\rm d} + \frac{3}{2} Y^{}_{\rm u} Y^\dagger_{\rm u}  Y^{}_{\rm u} Y^\dagger_{\rm u} 
	\nonumber
	\\
	&& + \left.\left. \frac{11}{4} Y^{}_{\rm d} Y^\dagger_{\rm d} Y^{}_{\rm d} Y^\dagger_{\rm d} - Y^{}_{\rm d} Y^\dagger_{\rm d} Y^{}_{\rm u} Y^\dagger_{\rm u} - \frac{1}{4} Y^{}_{\rm u} Y^\dagger_{\rm u} Y^{}_{\rm d} Y^\dagger_{\rm d} \right] \right\} Y^{}_{\rm u} \;,
	\nonumber
	\\
	\mu \frac{{\rm d} Y^{}_{\rm d}}{{\rm d} \mu} &=& \frac{1}{16\pi^2} \left\{ - \frac{5}{12} g^2_1 - \frac{9}{4}g^2_2 - 8g^2_3 + T - \frac{3}{2} \left( Y^{}_{\rm u} Y^\dagger_{\rm u} - Y^{}_{\rm d} Y^\dagger_{\rm d} \right)  + \frac{1}{16\pi^2} \left[ - \frac{127}{216} g^4_1 - \frac{23}{4} g^4_2 - 108 g^4_3  \right. \right.
	\nonumber
	\\
	&& - \frac{9}{4} g^2_1 g^2_2 + \frac{31}{9} g^2_1 g^2_3 + 9 g^2_2 g^2_3  + \frac{5}{24} g^2_1 {\rm Tr} \left( 15 Y^{}_l Y^\dagger_l + 17 Y^{}_{\rm u} Y^\dagger_{\rm u} + 5 Y^{}_{\rm d} Y^\dagger_{\rm d}  \right)  + \frac{15}{8} g^2_2 T  
	\nonumber
	\\
	&& + 20 g^2_3 {\rm Tr} \left( Y^{}_{\rm u} Y^\dagger_{\rm u}  + Y^{}_{\rm d} Y^\dagger_{\rm d} \right) + 6 \lambda^2  - \frac{9}{4} T^\prime +  \frac{3}{2} {\rm Tr} \left( Y^{}_{\rm u} Y^\dagger_{\rm u} Y^{}_{\rm d} Y^\dagger_{\rm d} \right) + \left( \frac{187}{48} g^2_1 + \frac{135}{16} g^2_2 + 16 g^2_3 \right.
	\nonumber
	\\
	&& - \left. 12\lambda - \frac{9}{4} T \right) Y^{}_{\rm d} Y^\dagger_{\rm d} + \left( -\frac{79}{48} g^2_1 + \frac{9}{16} g^2_2 - 16 g^2_3 + \frac{5}{4} T \right) Y^{}_{\rm u} Y^\dagger_{\rm u} + \frac{3}{2} Y^{}_{\rm d} Y^\dagger_{\rm d}  Y^{}_{\rm d} Y^\dagger_{\rm d} 
	\nonumber
	\\
	&& + \left.\left. \frac{11}{4} Y^{}_{\rm u} Y^\dagger_{\rm u} Y^{}_{\rm u} Y^\dagger_{\rm u}  - Y^{}_{\rm u} Y^\dagger_{\rm u} Y^{}_{\rm d} Y^\dagger_{\rm d} - \frac{1}{4} Y^{}_{\rm d} Y^\dagger_{\rm d} Y^{}_{\rm u} Y^\dagger_{\rm u} \right] \right\} Y^{}_{\rm d} \;.
\end{eqnarray}

\bibliographystyle{apsrev4-1}
\bibliography{biblio}

\end{document}